\documentclass[aps,prx,reprint,amsmath,amssymb,superscriptaddress,longbibliography,floatfix]{revtex4-2}
\usepackage{amsthm,mathtools,booktabs,graphicx,microtype,enumitem,bm}
\usepackage{xcolor}
\usepackage[colorlinks=true,linkcolor=blue!45!black,citecolor=blue!45!black,urlcolor=blue!55!black]{hyperref}
\newcommand{\Tr}{\operatorname{Tr}}
\newcommand{\id}{\mathcal{I}}
\newcommand{\Ad}{\operatorname{Ad}}
\newcommand{\Var}{\operatorname{Var}}
\newcommand{\norm}[1]{\left\lVert#1\right\rVert}
\newcommand{\cL}{\mathcal L}
\newcommand{\cD}{\mathcal D}
\newcommand{\us}{\,\mu\mathrm{s}}
\usepackage{physics}
\usepackage{cleveref}
\crefname{equation}{Eq.}{Eqs.}
\Crefname{equation}{Eq.}{Eqs.}
\crefname{algorithm}{algorithm}{algorithms}
\Crefname{algorithm}{Algorithm}{Algorithms}
\newtheorem{theorem}{Theorem}
\newtheorem{proposition}[theorem]{Proposition}

\newcounter{algorithm}

\newenvironment{algorithm}[1][]{\par\medskip\noindent\begin{minipage}{\columnwidth}\hrule\smallskip\small\renewcommand{\caption}[1]{\refstepcounter{algorithm}\textbf{Algorithm \thealgorithm. ##1}\par\smallskip}}{\smallskip\hrule\end{minipage}\par\medskip}
\setlist[enumerate]{itemsep=2pt,topsep=3pt,leftmargin=*}
\graphicspath{{./}}
\begin{document}
\title{Exploiting non-Markovian information backflow in quantum error mitigation}
\author{Xiantao Li}
\email{Xiantao.Li@psu.edu}
\affiliation{Department of Mathematics, Pennsylvania State University, University Park, Pennsylvania 16802, USA}
\date{\today}
\begin{abstract}

Quantum error mitigation reduces bias in noisy computations at the cost of additional measurements. For non-Markovian noise, however, when noise is inverted independently at every layer, this cost can include information lost temporarily and later returned by the environment. For invertible reduced dynamics, we use the recovery of distinguishability
to compare the sampling costs of layer-by-layer and end-of-circuit
inversion, with an exact cost ratio for scalar dephasing. 
For stationary Gaussian dephasing with fixed, commuting couplings, the bath correlation function (BCF) determines the reduced channel entirely even in the strong coupling regime. We translate the spectral condition for coherence trapping into a criterion for mitigation. The optimal terminal shot overhead for a Pauli observable is bounded uniformly in circuit duration if and only if the bath spectrum has a finite inverse-square frequency moment along every coherence contributing to that observable. 
For a more general bath initially uncorrelated with the register, we integrate the second-order response from the BCF before assigning sampling weights. We give two implementations of this second-order correction. For
Clifford circuits with Pauli couplings, the first estimates a target
Pauli observable by sampling Pauli measurements. The second samples
one of two terminal correction channels per shot and uses the same
channels and signed weights for every target observable.  With a controlled third-order Dyson remainder, both schemes have bias $O(|\lambda|^3)$ at fixed duration. The two-channel sampler has shot overhead $(1+2r_T)^2$, with $r_T$ bounded by the overlap of the bath spectrum and the circuit's control filters. For a time-independent ideal Hamiltonian and a bath of finite spectral weight, gaps around the coupled transition frequencies bound this shot overhead uniformly in duration. 

\end{abstract}
\maketitle

\section{Introduction}\label{sec:intro}
Noise is the ultimate limiting factor for circuit depth at which a quantum processor can resolve an observable of interest. Quantum error mitigation extends that effective depth by trading additional circuit executions for a reduction in the bias of the measured expectation value \cite{Temme,EndoReview,CaiReview,KimUtility2023}. In probabilistic error cancellation (PEC), one represents the inverse noise map as a signed combination of operations available on the device. Sampling those operations produces an unbiased estimate when the noise model and the operations are accurate. The sum of the absolute coefficients, denoted by $K$, determines the statistical cost of the inverse. At fixed precision the shot overhead scales as $K^2$.

The Markovian setting gives a natural starting point for understanding this cost. A continuous quantum dynamical semigroup on a finite register has a generator of Lindblad form \cite{GKS1976,Lindblad1976}. More generally, a differentiable, invertible, completely positive divisible evolution admits a time-local generator with nonnegative canonical rates \cite{Hall}. Each subsequent channel contracts the trace distance between possible input states or leaves it unchanged. For example, suppose each layer subjects a qubit to dephasing that
attenuates its $X$ expectation by a factor $0<q<1$. In the Heisenberg
picture, $\mathcal E_j^\dagger(X)=qX$. After $L$ layers, the measured
$X$ expectation is therefore $q^L$ times its ideal value. This example illustrates the mechanism behind exponential sampling bounds for persistent noise, including lower bounds that apply beyond PEC \cite{Takagi2022,Takagi,Quek2024}. This overhead can limit the precision attainable at a fixed shot budget in algorithms such as quantum phase estimation and amplitude-estimation-based quantum mean estimation, whose standard implementations require longer coherent circuits as the target precision increases.

When the environment retains memory across gates, even identifying the noise to be inverted becomes harder. A layer calibrated in isolation can begin with a different joint system--environment state from the same layer inside a circuit. Experiments on superconducting qubits have resolved such temporal correlations, including contributions from spectator degrees of freedom \cite{White,Agarwal,White2025,Oda,SchultzTimeCorrelated}. Remarkably, during a later interval, the environment can return information to the register and increase the distinguishability of states that an earlier interval brought closer together \cite{BLP2009,RHP}. Dynamical decoupling exploits temporal correlations by modulating the interaction during the circuit \cite{ViolaLloyd,Cywinski}. For fixed controls, how much of the accumulated noise must one invert to recover the final observable? This question is particularly important whenever the environment returns part of the signal before readout.

\begin{figure}[!t]
\centering\includegraphics[width=\columnwidth]{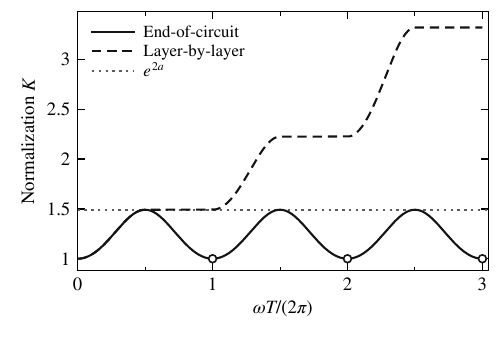}
\caption{PEC normalization for the harmonic Gaussian model with $a=0.2$. Open circles mark the full periods where $K_{\mathrm{end}}=1$ and the qubit can be measured directly. At odd half-periods it reaches $e^{2a}$. The layer-by-layer normalization accumulates the contraction in each period. Both curves use the same commuting $I/Z$ correction library, with the local curve evaluated in the fine-grid limit.  $K^2$ indicates the shot overhead.}\label{fig:extreme}
\end{figure}

This distinction can be illustrated by an analytically solvable model, with results shown in \Cref{fig:extreme}: A qubit is coupled longitudinally to a single Gaussian spectral line with coherence $e^{-\chi(t)}$, where $\chi(t)=a(1-\cos\omega t)$ and $a>0$. A simple mitigation strategy is to run the circuit and multiply its measured $X$ or $Y$ expectation by $e^{\chi(t)}$. At each full period $t=2\pi n/\omega$ the accumulated random phase vanishes for every realization. Information that left the qubit during the first half of the cycle has returned, the reduced evolution equals the ideal evolution, and direct measurement already gives the ideal expectation for every input state. Even at the times of greatest coherence loss, $T_n=(2n+1)\pi/\omega$, the terminal normalization stays fixed: $K_{\mathrm{end}}(T_n)=e^{2a}$. In contrast, inverting the noise independently through multiple layers containing each half-period extremum gives
\begin{equation}\label{eq:intro-harmonic}
  K_{\mathrm{layer}}(T_n)=e^{2a(n+1)}.
\end{equation}
What caused the exponential overhead is that the local inverse amplifies the signal in each interval of loss, then dephases it in each interval of recovery. This second operation costs no additional sampling overhead, but it preserves the cost already incurred in the first interval. 

For a single spectral line one can read the relevant times directly from the coherence. In a register with several noise couplings, a gate rotates the operators seen by the bath, and a given observable can depend on several coherences. Predicting the cost then requires both the bath correlation function (BCF) and the control history. We use these two inputs to connect the amount of information recovered by the register with the cost of a correction that can actually be sampled.

\subsection{Summary of results}
\emph{A benchmark for the cost of independent inversion.}
Building on results on channel inversion and distinguishability
\cite{Jiang,Regula2021,Chruscinski2011}, we first compare two ways of
mitigating the same invertible reduced evolution: inverting each noise
increment independently or inverting the accumulated noise once at
readout. Both procedures have access to all physical correction channels.
For a pair of register--reference states with the greatest relative loss
of distinguishability at readout, \Cref{eq:price-endpoint} relates the
two costs: each recovery along that pair's trajectory contributes to a
lower bound on the shot-overhead ratio. Scalar dephasing attains the
bound. For a coherence factor $e^{-\chi(t)}$ with absolutely continuous
$\chi$, the fine-time-step limit gives
$\log(K_{\mathrm{layer}}/K_{\mathrm{end}})
=\int_0^T[-\dot\chi(t)]_+\,dt$.
Thus the coherence history determines exactly how much independent
inversion pays for losses that are later reversed.

\emph{Physical errors with bounded mitigation cost.}
For commuting Gaussian dephasing, Theorem~\ref{thm:spectral} shows explicitly that the exact shot overhead for a Pauli observable stays bounded for all durations if and only if $\int j_u(\omega)\omega^{-2}\,d\omega$ is finite for every coherence $u$ contributing to that observable. Table~\ref{tab:spectral-scaling} translates the low-frequency behavior into a cost. For example, when the dephasing noise is concentrated at a single
nonzero frequency, the terminal shot overhead remains bounded at all
times and returns to one after each full period as the coherence
recovers.

\emph{BCF-based error mitigation.}
For general noise, we integrate the second-order bath response along the ideal circuit. Positive and negative rates in the same operator direction can cancel before sampling weights are assigned, as seen in the matrix $D(T)$ of \Cref{eq:signed-general-dissipation}. For Clifford circuits with Pauli couplings, Algorithm~\ref{alg:observable-compiler}
compiles the correction for a target Pauli observable into a weighted
sum of Pauli measurements. The second implementation samples one of
two terminal channels per shot and uses the same channels and signed
weights for every observable, with shot overhead $(1+2r_T)^2$. For both constructions, Theorem~\ref{thm:cumulative-cost} gives an $O(|\lambda|^3)$ fixed-duration bias under its Dyson remainder assumption. For a time-independent ideal Hamiltonian, Proposition~\ref{prop:shifted-moments} gives a sufficient spectral condition for the two-channel sampling cost to remain bounded even for noncommuting couplings. Overall, these results give a noise-aware prescription for QEM and identify
device noise that can be inexpensive to mitigate. Even a large loss
of coherence during a circuit can carry little sampling cost if the
signal recovers before readout. The bath spectrum, together with the
prescribed controls and target observable, provides a way to assess
when a device can benefit from this recovery.

\subsection{Relation to previous work}
A microscopic description of the noise gives a natural starting point
for estimating the cost of mitigation. Wang and Li~\cite{WangLi}
derived correction operations from bath correlation functions and
bounded the cost of applying them stepwise. Their bound takes the
absolute magnitude of the correlations at each delay, so contributions
of opposite sign cannot cancel in the cost estimate. Why the sign
matters can already be seen in the earlier rate-based analysis of
Hakoshima et al.~\cite{Hakoshima}. In their dephasing model, a positive
decay rate increases the stochastic PEC normalization, while a negative
rate adds no further cost. Even when commuting corrections are moved
to the output, independent sampling retains the cost of the preceding
decay. Endo, Hakoshima, and Shitara~\cite{Endo} gave a complementary
interpretation in terms of distinguishability: information returned
during logical feedback lowers a sampling bound for mitigating the
remaining logical noise. These observations motivate the question
studied here. If the circuit and readout time are fixed, how much of
the earlier sampling cost can be avoided by combining the corrections
before drawing them?

A bath-dependent inverse can be constructed explicitly for particular
gates, as Ahn and Park~\cite{AhnPark} demonstrated for a CNOT gate
coupled to an oscillator bath. For a circuit containing many gates,
the choice of what to invert becomes important. The original PEC
formulation already samples a signed ensemble of complete
circuits~\cite{Temme}. Its weights can be obtained from characterized
imperfect operations~\cite{EndoPractical} or learned from training
circuits, including in the presence of temporal
correlations~\cite{Strikis2021}. Blockwise mitigation and quasiprobability
optimization show that combining operations before sampling can reduce
the coefficient norm~\cite{Compressed,Mosaic,PiveteauQPD,Scheiber2025}.
When only a final expectation is required, the correction can also be
carried into classical postprocessing by tensor-network
mitigation~\cite{FilippovTEM,FilippovScalability,FischerTEMHardware},
or into the choice of measured observable~\cite{Watanabe,Zhao}.
For a specified invertible endpoint channel, the implementability
measure of Jiang, Wang, and Wang~\cite{Jiang} gives the minimum
full-channel normalization over all CPTP corrections,
$K_{\mathrm{end}}=2^{\nu(\widetilde\Phi_T^{-1})}$.
This characterization applies even when the channel is generated by
a bath with memory. Observable propagation~\cite{Qin,Eddins} and
calibrated rescaling~\cite{Urbanek2021,Vovrosh2021} offer ways to
recover a particular expectation at readout.

Liu, Xiao, and Cai represent a process with memory by a Choi channel
with one register for each noise interval. They decompose its inverse
into products of physical operations and sample the corresponding
operations at successive times with the same signed coefficients
\cite[Sec.~IV, Eqs.~(8) and (9)]{Liu}. A single decomposition then
corrects the process for any gate between the two noise intervals.
Our harmonic example shows how fixing that gate changes the task.
Over two successive half-periods, the minimum normalization for
inverting the full Choi channel is $K_{\mathrm{process}}=e^{8a}$.
With the identity between the intervals, the second half-period restores
the initial coherence. The two independent inverse factors have
$K_{\mathrm{layer}}=e^{2a}$, but all four branches of the Choi-channel
decomposition implement the same circuit. Combining their coefficients
before sampling gives $K_{\mathrm{end}}=1$
(Appendix~\ref{app:liu-harmonic}).

For invertible reduced dynamics,
the accumulated logarithmic recovery of a weighted register--reference
state pair attaining the worst final loss lower-bounds
$\log(K_{\mathrm{layer}}/K_{\mathrm{end}})$. For scalar dephasing the
relation is exact, as shown in \Cref{eq:log-backflow}. To turn this
comparison into a mitigation procedure, we integrate the bath
correlations along the control history before assigning sampling
weights. For commuting Gaussian dephasing, this gives an exact
spectral criterion for the optimal shot overhead of a target Pauli
observable. For more general  baths, it gives two terminal
corrections with fixed-duration bias bounds under a controlled
second-order expansion. The required spectral information can be
probed through single-qubit noise spectroscopy and measurements of
cross-correlations between qubits~\cite{Bylander2011,VonLupke2020}.

The rest of the paper is organized as follows. Section~\ref{sec:price} compares layer-by-layer and end-of-circuit inversion and relates their sampling costs to the recovery of distinguishability. Section~\ref{sec:spectral} uses this comparison to characterize the bath spectra for which commuting Gaussian dephasing admits bounded mitigation overhead. Section~\ref{sec:control} develops a second-order correction for more general baths and bounds the bias in the final observable. Section~\ref{sec:numerics} tests these predictions in numerical models, and Sec.~\ref{sec:discussion} discusses their implications for noise characterization and mitigation on quantum devices.

\section{Sampling cost and recovery of information}\label{sec:price}

\subsection{General PEC inversion}
Throughout this paper, we use a continuous-time description of the register dynamics to track how information is lost and recovered over the course of a circuit. To compare layer-by-layer and end-of-circuit mitigation, we evaluate this evolution at the layer boundaries and at the final readout time. Let $\Phi_t$ denote the physical reduced channel of the register from time $0$ to $t$, and let $\mathcal U_t=\Ad_{U_t}$ denote the corresponding ideal evolution, with $\Ad_{U_t}(X)=U_t X U_t^\dagger$. In the interaction frame of the ideal circuit, the accumulated noise is described by $\widetilde\Phi_t=\mathcal U_t^{-1}\circ\Phi_t$. In the absence of noise, $\widetilde\Phi_t=\id$, where $\id(X)=X$ denotes the identity superoperator.

Probabilistic error cancellation (PEC) represents the inverse of a noise channel $\mathcal N$ as a signed combination of physical channels $\mathcal B_i$,
\begin{equation}\label{eq:pec-intro}
 \mathcal N^{-1}=\sum_i q_i\mathcal B_i,\qquad
 K=\sum_i|q_i|.
\end{equation}
On each shot, one draws $i$ with probability $|q_i|/K$, applies
$\mathcal B_i$, and records the weighted outcome
$K\operatorname{sgn}(q_i)y$, where $y$ is the measured outcome and
$|y|\leq 1$. The sample mean of these weighted outcomes is an unbiased
estimator of the ideal expectation value~\cite{Temme}.  Estimating this value to statistical precision $\epsilon$ with failure probability at most $\delta$ requires $O(K^2\epsilon^{-2}\log(2/\delta))$ independent shots. We therefore use $K$ for the normalization and $K^2$ for the shot overhead throughout the paper.

\subsection{Independent inverses and recovered distinguishability}
\label{sec:gross-net}

We begin with a comparison that can be made directly from the reduced
dynamics. Following the channel-cost approach of Jiang, Wang, and
Wang~\cite{Jiang}, we compare two ways of inverting the noise of a
prescribed circuit. We then use the distinguishability of two preparations
to identify a contribution to their cost difference.

Choose times $0=t_0<\cdots<t_L=T$ at gate or layer boundaries, or within
a gate or idle period when finer time resolution is needed. Let
$\Phi_j=\widetilde\Phi_{t_j}$ be the cumulative interaction-frame noise
channel and assume each $\Phi_j$ is invertible, with $\Phi_0=\mathcal I$.
Layer-by-layer PEC samples a decomposition of each inverse factor
$\mathcal F_j=\Phi_{j-1}\Phi_j^{-1}$ independently. End-of-circuit PEC
samples one decomposition of their product
$\mathcal F_1\cdots\mathcal F_L=\Phi_L^{-1}$. Both procedures can be
implemented at the output. For independent sampling, the factors are
applied in reverse chronological order in the output frame.

For a Hermiticity-preserving, trace-preserving map, the minimum
normalization over all CPTP correction channels equals its diamond
norm~\cite[Theorem~3]{Regula2021}. The two minimum normalizations are
therefore
\begin{equation}\label{eq:optimal-costs}
 K_{\mathrm{layer}}=\prod_{j=1}^L\|\mathcal F_j\|_\diamond,\qquad
 K_{\mathrm{end}}=\|\Phi_L^{-1}\|_\diamond .
\end{equation}
The composition bound $K_{\mathrm{end}}\le K_{\mathrm{layer}}$ is the
subadditivity of the logarithmic implementation cost proved
in Ref.~\cite[Theorem~7]{Jiang}.

To follow the loss and recovery of distinguishability, we use the
trace-distance approach of Breuer, Laine, and Piilo~\cite{BLP2009} in
the register--reference formulation of Chru\'sci\'nski, Kossakowski,
and Rivas~\cite{Chruscinski2011}. Let $\rho_0,\rho_1$ be two possible
preparations of the register and an isolated reference of the same
dimension, with prior probabilities $p$ and $1-p$. Write their weighted
difference as $Z=p\rho_0-(1-p)\rho_1\ne0$. The optimal probability of
identifying the preparation at $t_j$ is $(1+d_j(Z))/2$, with
\begin{equation}\label{eq:price-path}
 d_j(Z)=\|(\Phi_j\otimes\mathcal I_R)(Z)\|_1,\qquad
 c_j(Z)=\log\frac{d_{j-1}(Z)}{d_j(Z)} .
\end{equation}
Thus $c_j>0$ records a loss of distinguishability and $c_j<0$ records a
recovery. We use logarithmic changes because independent PEC
normalizations multiply. A logarithmic coherence ratio also appears
in the logical dephasing example of Endo et al.~\cite[Eq.~(5)]{Endo}.

The inverse factor $\mathcal F_j$ takes the evolved difference at $t_j$
back to that at $t_{j-1}$. Its diamond norm is at least
$d_{j-1}(Z)/d_j(Z)$ and, by trace preservation, at least one. Taking the
product gives $\log K_{\mathrm{layer}}\ge\sum_j[c_j(Z)]_+$, where
$[x]_+=\max(x,0)$. Now choose a pair $Z_*$ attaining the greatest net
loss, $d_0(Z_*)/d_L(Z_*)=K_{\mathrm{end}}$. Such a pair exists by
invertibility and the Hermitian-input characterization of the diamond
norm~\cite{Regula2021}. For this same pair the sum of the $c_j$ telescopes
to $\log K_{\mathrm{end}}$. Writing
$\mathcal I_{\mathrm{rec}}(Z)=\sum_j[-c_j(Z)]_+$ for its accumulated
logarithmic recovery gives the shot-overhead comparison
\begin{equation}\label{eq:price-endpoint}
 \frac{K_{\mathrm{layer}}^2}{K_{\mathrm{end}}^2}
 \ge \exp\!\left[2\mathcal I_{\mathrm{rec}}(Z_*)\right].
\end{equation}
For CP-divisible dynamics, each forward increment is CPTP, so
$d_j(Z)\le d_{j-1}(Z)$ for every pair~\cite{Chruscinski2011}.
The recovery term then vanishes, although the cost ratio can still
exceed one, as for two successive qubit depolarizing channels~\cite{Jiang}.

Scalar dephasing attains \Cref{eq:price-endpoint}, as shown in
\Cref{eq:log-backflow}. To turn this comparison into a mitigation
procedure, we need to determine what error remains at readout.
Section~\ref{sec:spectral} uses bath correlations to do this exactly for
commuting Gaussian dephasing. Section~\ref{sec:control} uses the same
input to construct corrections through second order for general centered
baths.

\section{Exact mitigation for commuting Gaussian dephasing}\label{sec:spectral}\label{sec:model}
Ramsey and echo measurements on superconducting qubits have been described by Gaussian frequency noise \cite{Bylander2011}, while engineered NMR environments exhibit coherence loss and recovery under dephasing \cite{Khurana2019}. Two-qubit spectroscopy also gives access to cross-correlations between the local fields \cite{VonLupke2020}. These experiments motivate a question we can answer exactly for commuting Gaussian noise. Which part of the measured spectrum determines the cost of recovering a given observable?

\subsection{Bath correlations and the observable}
Consider an $n$-qubit register of dimension $d=2^n$, coupled to a stationary bath by the Hamiltonian \cite{breuer2002theory}
\begin{equation}\label{eq:hamiltonian}
 \begin{aligned}
 H_{\rm tot}(t)&=H_{\rm id}(t)+H_B
+\lambda\sum_{\alpha=1}^{m}A_\alpha\otimes B_\alpha,\\
 \rho_{SB}(0)&=\rho_0\otimes\tau_B.
 \end{aligned}
\end{equation}
We take $A_\alpha$ and $B_\alpha$ to be Hermitian. Each system coupling $A_\alpha$ may act on one or more qubits. We include all bath-only terms in $H_B$, take the remaining system couplings to be traceless, and assume $[H_B,\tau_B]=0$ for this choice. Centering the coupling, $\Tr(\tau_BB_\alpha)=0$, removes the term linear in $\lambda$ from the reduced evolution \cite{breuer2002theory}.  This centering step is assumed throughout the paper.

If $U_t$ is the ideal propagator, set $\widetilde A_\alpha(t)=U_t^\dagger A_\alpha U_t$, $\mathcal U_t=\Ad_{U_t}$, and $B_\alpha(t)=e^{iH_Bt}B_\alpha e^{-iH_Bt}$. The two-point bath correlation function (BCF) is
\begin{equation}\label{eq:bcf}
 C_{\alpha\beta}(t-s)=\Tr[\tau_BB_\alpha(t)B_\beta(s)].
\end{equation}
Cross-correlations allow spatially correlated noise even when individual couplings are local. Section~\ref{sec:control} uses this BCF without Gaussian or commuting assumptions.

\subsection{Commuting couplings and the exact influence map}\label{sec:exact-gaussian}

Let $U_t$ denote the ideal register evolution. We assume for now that it
preserves each system coupling operator,
$\widetilde A_\alpha(t)=U_t^\dagger A_\alpha U_t=A_\alpha$
for $0\leq t\leq T$. We also assume that the Hermitian operators
$A_\alpha$ commute with one another. This allows us to choose a
common register eigenbasis $\{\ket z\}$ and perform explicit calculations. Toward this end, we write
$A_\alpha\ket z=a_\alpha(z)\ket z$.
For the exact attenuation formula, take the bath to be a centered real stationary Gaussian field $\xi_\alpha(t)$ with covariance $C_{\alpha\beta}(t-s)=\mathbb E[\xi_\alpha(t)\xi_\beta(s)]$. Set $\theta_\alpha(T)=\lambda\int_0^T\xi_\alpha(t)dt$ and $\operatorname{ad}_A(X)=[A,X]$. The Gaussian influence map is
\begin{align}
 G(T)&=\lambda^2\int_0^Tdt\int_0^Tds\,C(t-s)\succeq0,\label{eq:covariance}\\
 \widetilde\Phi_T
 &=\mathbb E\exp\!\left[-i\sum_\alpha\theta_\alpha(T)\operatorname{ad}_{A_\alpha}\right]
 \notag\\
 &=\exp\!\left[-\frac12\sum_{\alpha\beta}G_{\alpha\beta}(T)
       \operatorname{ad}_{A_\alpha}\operatorname{ad}_{A_\beta}\right].
 \label{eq:gaussian-influence}
\end{align}
The commuting generators allow us to average the accumulated phase using the Gaussian characteristic function. Its logarithm is quadratic, which makes the influence functional in \Cref{eq:gaussian-influence} \emph{exact} at every duration \cite{Doll,Hangleiter2021,HangleiterErratum}.

The phase difference between a pair of common eigenstates involves only the differences $d_{zz'}=a(z)-a(z')$ of their coupling eigenvalues. With the covariance in \Cref{eq:covariance}, this gives
\begin{equation}\label{eq:schur}
 \begin{aligned}
 (\widetilde\Phi_T\rho)_{zz'}&=k_T(z,z')\rho_{zz'},\\
 k_T(z,z')&=e^{-\chi_{zz'}(T)},\\
 \chi_{zz'}(T)&=\tfrac12d_{zz'}^\dag G(T)d_{zz'}.
 \end{aligned}
\end{equation}

We thus obtain in closed form the cumulative noise channel $\Phi_L=\widetilde\Phi_T$ used in \Cref{eq:optimal-costs}. Each coherence
$\ket z\!\bra{z'}$ is an eigenoperator of this channel, with eigenvalue
$k_T(z,z')$ given in \Cref{eq:schur}. The physical register evolution
is $\Phi_T=\mathcal U_T\circ\widetilde\Phi_T$. To correct a terminal
observable, we can therefore invert the factors on the coherences
that contribute to its expectation.

\subsection{The spectrum-dependent sampling overhead}

We now use the channel in \Cref{eq:schur} to recover the expectation
of a single observable by changing the final measurement. Let $O$ be
the desired output observable and suppose that its interaction-frame
form $P=U_T^\dagger O U_T$ is a Pauli string in the computational
basis $\{\ket z\}$ in which the couplings are diagonal.

If $P$ is
diagonal, then $\widetilde\Phi_T^\dagger(P)=P$, and the original
measurement of $O$ already gives the ideal expectation.
Otherwise, let $\mathcal J_P$ be the set of qubits on which $P$ acts
as $X$ or $Y$. Acting on $\ket z$, the Pauli string flips precisely
these bits and multiplies the resulting state $\ket{z'}$ by a phase.
This partitions the computational basis into disjoint pairs
$u=\{z,z'\}$. Only the coherence within each pair contributes to
the expectation of $P$.

To correct the measurement, we will choose an ordering for each pair and write
$k_u=k_T(z,z')$ and $d_u=a(z)-a(z')$. By \Cref{eq:schur}, the adjoint
noise channel multiplies the $(z,z')$ entry of an observable by
$\overline{k_u}$. We therefore obtain the corrected observable $O_T$
by dividing $P_{zz'}$ by $\overline{k_u}$ on each pair and setting the
reverse entry to its complex conjugate. This gives
$\widetilde\Phi_T^\dagger(O_T)=P$. On each shot, we measure
$U_T O_T U_T^\dagger$ on the physical output and record its eigenvalue.
For any input state $\rho$, the mean score is
$\Tr[O_T\widetilde\Phi_T(\rho)]=\Tr[P\rho]$, i.e., the desired ideal
expectation.

Each pair contributes a two-dimensional block of $O_T$ with zero
diagonal and off-diagonal entry $P_{zz'}/\overline{k_u}$. Since
$|P_{zz'}|=1$, the two possible scores in this block are
$\pm |k_u|^{-1}$. The most strongly attenuated coherence contributing
to $P$ thus determines the score bound $K_P(T)$ and the corresponding
worst-case shot overhead $K_P(T)^2$,

\begin{equation}\label{eq:readout-radius-main}
 \begin{aligned}
 K_P(T)&=\| O_T\|_\infty=\max_u|k_u|^{-1},\\
 K_P(T)^2&=\exp\!\left[\max_u d_u^\dag G(T)d_u\right].
 \end{aligned}
\end{equation}
In 
Appendix~\ref{app:readout-proof} we prove that every terminal single-copy
estimator unbiased for all inputs must have a score radius at least
$K_P(T)$.

The noise spectrum is a matrix-valued density $J(\omega)\succeq0$, which is assumed to be continuous away from $\omega=0$ with finite total weight $\int_{\mathbb R}\Tr J(\omega)d\omega$. It has the properties that 
\begin{align}
 C(\tau)&=\int_{\mathbb R}e^{-i\omega\tau}J(\omega)\,d\omega,
 \qquad J(-\omega)=\overline{J(\omega)},
 \label{eq:spectral-measure}\\
 G(T)&=\lambda^2\int_{\mathbb R}|F_T(\omega)|^2J(\omega)\,d\omega,
 \notag\\
 F_T(\omega)&=\int_0^T e^{i\omega t}dt,\qquad
 |F_T(\omega)|^2=\frac{4\sin^2(\omega T/2)}{\omega^2}.
 \label{eq:spectral-covariance}
\end{align}
Here $F_T(0)=T$. This is the usual Ramsey filter applied to the measured spectrum \cite{Cywinski,Bylander2011}. For each target pair, set $j_u(\omega)=\lambda^2d_u^\dag J(\omega)d_u\ge0$. Combining \Cref{eq:readout-radius-main,eq:spectral-covariance} gives the overhead $K_P(T)^2=\exp[\max_u\int|F_T(\omega)|^2j_u(\omega)d\omega]$.

\begin{theorem}[Spectral criterion and explicit shot overhead]
\label{thm:spectral}
Under the stationary commuting Gaussian assumptions above, define
\begin{equation}\label{eq:spectral-moment}
 \mathfrak m_u=\int_{\mathbb R}\frac{j_u(\omega)}{\omega^2}\,d\omega.
\end{equation}
For a fixed finite-dimensional target, the exact overhead $K_P(T)^2$ is uniformly bounded in $T$ if and only if every $\mathfrak m_u$ is finite. Specifically,
\begin{equation}\label{eq:spectral-iff}
 \begin{aligned}
 \exp[2\max_u\mathfrak m_u]&\leq\sup_{T\ge0}K_P(T)^2
 \leq\exp[4\max_u\mathfrak m_u].
 \end{aligned}
\end{equation}
For a density with finite inverse-square moments, the limiting overhead is
\begin{equation}\label{eq:spectral-density-limit}
 \lim_{T\to\infty}K_P(T)^2=\exp[2\max_u\mathfrak m_u].
\end{equation}

\end{theorem}

Since $\int_{|\omega|\ge\omega_0}j_u(\omega)\omega^{-2}d\omega\le\omega_0^{-2}\int j_u(\omega)d\omega$, $\forall \omega_0>0$,  the criterion therefore depends only on spectral weight near zero. Table~\ref{tab:spectral-scaling} translates this dependence into shot overhead for several bath asymptotics. For a scalar continuum spectrum, the inverse-square moment is familiar from coherence trapping~\cite{Addis,Trushechkin}. \Cref{eq:spectral-iff} takes the maximum over the coherences seen by the target and gives sharp constants for the optimal overhead. Its uniform criterion also covers discrete lines whose coherences oscillate indefinitely. The methods from Section~\ref{sec:price} determine how much independent sampling pays for coherence that returns along the way.

As a concrete example, consider a thermal bosonic bath with linear coupling. If $j_{u,\mathrm{bare}}$ is the one-sided bare spectrum in a target direction, its two-sided symmetrized spectrum, which determines the attenuation, is
\begin{equation}\label{eq:thermal-spectrum}
 \begin{aligned}
 j_u(\pm\omega)&=\tfrac12j_{u,\mathrm{bare}}(\omega)
                  [2n_B(\omega)+1],\quad \omega>0,\\
 n_B(\omega)&=(e^{\beta\omega}-1)^{-1}.
 \end{aligned}
\end{equation}
The occupation factor is $2n_B+1=\coth(\beta\omega/2)$. Consequently, a bare power law $j_{u,\mathrm{bare}}(\omega)\asymp\omega^s$ gives $p=s-1$ in Table~\ref{tab:spectral-scaling} at positive temperature and $p=s$ at zero temperature. Bounded cost requires $s>2$ and $s>1$, respectively, with an integrable high-frequency cutoff \cite{Trushechkin}. Thermal occupation thus shifts the threshold by one power of frequency even for a super-Ohmic bath.

\begin{table}[t]
\centering\small
\begin{tabular}{@{}p{0.47\columnwidth}p{0.47\columnwidth}@{}}\toprule
Target spectrum near zero & Terminal shot-cost scaling $K_P(T)^2$\\\midrule
$j_*\asymp|\omega|^p$, $-1<p<1$ & $\exp[\Theta(T^{1-p})]$\\
$j_*\asymp|\omega|$, $p=1$ & $\exp[\Theta(\log T)]$ (polynomial)\\
$j_*\asymp|\omega|^p$, $p>1$, or a gap & $O(1)$\\
Zero-frequency spectral line & $\exp[\Theta(T^2)]$\\
\bottomrule
\end{tabular}
\caption{Low-frequency spectrum and terminal sampling cost for commuting Gaussian dephasing. Power-law rows describe $j_*(\omega)=\max_u j_u(\omega)$, with two-sided comparability near zero and finite covariance. Stationary $1/f$ noise requires an infrared cutoff. Appendix~\ref{app:filter-scope} gives the bounds.}
\label{tab:spectral-scaling}
\end{table}

The density result also holds across finite jumps in $J$. A discrete line with projected weight $w$ at $\omega_\ell\ne0$ contributes $w/\omega_\ell^2$ to $\mathfrak m_u$, while a positive weight at zero makes the moment infinite. Lines can sustain oscillations in the overhead, as the following example illustrates. In Appendix~\ref{app:spectral-proof} we prove the boundedness criterion for general spectral measures.

\smallskip 
For a register initially uncorrelated with a linearly coupled Gaussian
bath, the two-point bath correlations and the prescribed ideal evolution
also determine $\widetilde\Phi_T$ when the interaction-frame coupling
operators do not commute, see e.g. \cite{breuer2004time}. Huang et al.\ give an exact expansion of this
map in which the bath correlations weight time-ordered products of
system superoperators~\cite[Theorem~2.7 and Eq.~(2.28)]{HuangZhuParkLin}.
When the resulting map is invertible, its inverse can be decomposed at the terminal time using \Cref{eq:pec-intro}. A natural future direction is how tightly the bath spectrum and the circuit controls can bound the shot overhead of exact terminal PEC when the system coupling operators do not commute.

\subsection{Analytical example: repeated recovery at bounded terminal cost} 
We now consider an exactly solvable environment, with 
a qubit with coupling $Z/2$ in a stationary Gaussian field
\begin{equation}\label{eq:harmonic-bath}
 \begin{aligned}
 \xi(t)&=x\cos\omega t+y\sin\omega t,
 x,y\overset{\rm iid}{\sim}\mathcal N(0,\sigma^2), \\ 
 C(\tau)&=\sigma^2\cos\omega\tau.
 \end{aligned}
\end{equation}
This Gaussian field has two spectral lines, each of weight $\sigma^2/2$, at $\pm\omega$. With $a=\lambda^2\sigma^2/\omega^2$,
\begin{equation}\label{eq:harmonic-rate}
 \begin{aligned}
 \chi(T)&=a(1-\cos\omega T),\\
 \dot\rho&=\gamma(t)(Z\rho Z-\rho),\qquad
 \gamma(t)=\tfrac12a\omega\sin\omega t.
 \end{aligned}
\end{equation}
At $T=2n\pi/\omega$ the integrated field vanishes for every realization of $x,y$. The qubit has returned to its ideal state, so measuring the original observable already gives the ideal result. At intermediate times one can recover an $X$ or $Y$ expectation by multiplying its noisy value by $e^{\chi(T)}$. 
For either transverse observable, the two eigenvalues of $Z/2$ differ
by one, so the projected spectrum is
\[
 j(\nu)=\frac{\lambda^2\sigma^2}{2}
 \bigl[\delta(\nu-\omega)+\delta(\nu+\omega)\bigr].
\]
Its inverse-square moment in \Cref{eq:spectral-moment} is
$\mathfrak m=\lambda^2\sigma^2/\omega^2=a$. The spectral bound
\Cref{eq:spectral-iff} therefore gives $K_{\rm end}^2\le e^{4a}$
for every $T$, with equality at the odd half-periods
$T=(2n+1)\pi/\omega$, where $\chi(T)=2a$.

In contrast, to see what the local sampler pays, consider a layer grid containing each half-period endpoint. At $T_n=(2n+1)\pi/\omega$ the qubit has completed $n+1$ intervals of coherence loss, each with $\Delta\chi=2a$, giving $K_{\rm layer}^2=e^{4a(n+1)}$. Between successive losses the bath returns coherence. The local inverse then dephases the returned coherence after having paid to restore it in the preceding interval.

The example also explains why low-frequency spectral weight matters. Broadening the line to $C(\tau)=\sigma^2e^{-\kappa|\tau|}\cos\omega\tau$, $\kappa>0$, gives
\begin{equation}\label{eq:damped-asymptotic}
 \begin{aligned}
 \chi(T)&=\lambda^2\sigma^2\operatorname{Re}
 \left[\frac{T}{z}-\frac{1-e^{-zT}}{z^2}\right]\\
 &=\frac{\lambda^2\sigma^2\kappa}{\kappa^2+\omega^2}T+O(1),
 \qquad z=\kappa-i\omega.
 \end{aligned}
\end{equation}

Consequently, broadening the spectral line introduces weight at zero frequency, so the
terminal shot overhead now grows with circuit duration. Its exponential
growth rate follows directly from \Cref{eq:damped-asymptotic}:
\[
 \lim_{T\to\infty}\frac{\log K_{\rm end}^2}{T}
 =\lim_{T\to\infty}\frac{2\chi(T)}{T}
 =\frac{2\lambda^2\sigma^2\kappa}{\kappa^2+\omega^2}.
\]
The oscillations of the bath correlation function still matter.
Replacing $C(\tau)$ by its absolute envelope
$\sigma^2 e^{-\kappa|\tau|}$, as in an absolute-BCF
estimate~\cite{WangLi}, gives the upper bound
$2\lambda^2\sigma^2/\kappa$ on this rate. For a narrow line
with $\kappa\ll\omega$, the exact rate is smaller than this bound
by a factor of $1+(\omega/\kappa)^2$.

\section{Beyond Gaussian dephasing: corrections from bath correlations}\label{sec:control}
We now allow a non-Gaussian bath and noncommuting interaction-frame couplings. For a  stationary bath and the factorized initial state in \Cref{eq:hamiltonian}, the ordered two-point BCF in \Cref{eq:bcf} determines the leading reduced error \cite{breuer2004time}. We integrate that error over the prescribed circuit before constructing its inverse. A contribution from an interval of decoherence can then cancel against a later contribution before either is assigned a sampling cost.

We give two implementations of this same second-order inverse. For a single target observable and Clifford circuits with Pauli couplings, Algorithm~\ref{alg:observable-compiler} compiles the correction into Pauli measurements on the unmodified noisy output. When terminal channels can be synthesized, the correction in \Cref{eq:two-branch-correction} samples two such channels using weights shared by all observables. \Cref{thm:cumulative-cost} gives the sampling cost and bias of each implementation.

\subsection{Noise integration}
Let $\Phi_T$ be the actual reduced map without mitigation and
$\mathcal U_T=\Ad_{U_T}$ the ideal unitary evolution. From the same input,
\begin{equation}
\begin{aligned} \rho_{\rm id}(T)&=\mathcal U_T(\rho_0),\\ \rho_{\rm noisy}(T)&=\Phi_T(\rho_0),\qquad \widetilde\Phi_T=\mathcal U_T^{-1}\Phi_T.\end{aligned}
\end{equation}
The ideal interaction-picture map is the identity. Put
$V(t)=\sum_\alpha\widetilde A_\alpha(t)\otimes B_\alpha(t)$.
The joint Dyson expansion, followed by the bath trace, has no first-order
term because the bath is centered \cite{breuer2002theory,breuer2004time}. At second order the derivative of the reduced Dyson coefficient is the time-convolutionless generator $\cL_2(t)$, which yields
\begin{align}
  \mathcal K_2(T)&=\int_0^T\cL_2(t)dt,\nonumber\\
 \widetilde\Phi_T&=\id+\mathcal K_2(T)+\mathcal R_3(T),\label{eq:cumulative}\\
 \cL_2(t)X&=-\lambda^2\int_0^t\Tr_B[V(t),[V(s),X\otimes\tau_B]]ds.\nonumber
\end{align}
At fixed $T$, $\|\mathcal R_3(T)\|_\diamond=O(|\lambda|^3)$ under a
controlled Dyson expansion. The remainder is $O(\lambda^4)$ when the third reduced coefficient vanishes and the fourth-order remainder is controlled \cite{breuer2002theory}. Centered Gaussian fields and symmetric stationary telegraph noise both have vanishing odd moments. 

To expose the coherent and dissipative contributions, we follow \cite{Hall} and choose a fixed
traceless Hermitian basis $\Tr(F_aF_b)=\delta_{ab}$ and expand the traceless
couplings as $\widetilde A_\alpha(t)=\sum_a a_{a\alpha}(t)F_a$.
The time ordering in the double commutator defines a matrix $ Q(t)$. Its Hermitian and anti-Hermitian parts give the dissipative rates and Lamb shift, respectively, through
\begin{align}
  Q_{ab}(t)&=\lambda^2\sum_{\alpha\beta}\int_0^t
 C_{\alpha\beta}(t-s)a_{a\beta}(s)a_{b\alpha}(t)\,ds,
 \label{eq:qmatrix}\\
 \Gamma_2^{\rm can}(t)&= Q(t)+ Q(t)^\dagger,\qquad
 S(t)=\frac{ Q(t)- Q(t)^\dagger}{2i},\nonumber\\
 H_{\rm LS}(t)&=\sum_{ab}S_{ab}(t)F_bF_a,
 \label{eq:lamb}
\end{align}
so that
\begin{equation}\label{eq:tcl-basis}
\begin{aligned}
 \cL_2(t)X={}&-i[H_{\rm LS}(t),X]\\
 &+\sum_{ab}[\Gamma_2^{\rm can}(t)]_{ab}
 \left(F_aXF_b-\tfrac12\{F_bF_a,X\}\right).
\end{aligned}
\end{equation}
The eigenvalues of $\Gamma_2^{\rm can}(t)$ are the second-order canonical rates. Both the matrix and the Lamb shift are determined by the BCF and the ideal control history. The integrated Lamb shift can be compensated to this order by a terminal unitary rotation with normalization $K=1$.

For $B=U_T^\dagger O U_T$, the leading observable error is
$\Tr[\rho_0\mathcal K_2(T)^\dagger B]$. We therefore combine the signed
coefficients of \Cref{eq:tcl-basis} over the complete evolution before
constructing the inverse.

\subsection{Signed cumulative dissipation}
The ordered BCF in \Cref{eq:bcf} satisfies
$\overline{C_{\alpha\beta}(t-s)}=C_{\beta\alpha}(s-t)$, and the
coefficients $a_{a\alpha}(t)$ are real in the chosen Hermitian basis.
Using this symmetry in \Cref{eq:qmatrix}, we can exchange the two
integration times and relabel the bath indices in the adjoint
contribution. The integrals of $Q(t)$ and $Q(t)^\dagger$ then have
the same integrand on complementary halves of $[0,T]^2$. Their sum
gives the full-square integral for the cumulative dissipative
coefficient,
\begin{align}
 D_{ab}(T)&=\lambda^2\sum_{\alpha\beta}\int_0^Tdt\int_0^Tds\,
 a_{a\alpha}(t)a_{b\beta}(s)C_{\beta\alpha}(s-t),\label{eq:cumulative-gram}\\
 H_2(T)&=\int_0^T H_{\rm LS}(t)dt.
\end{align}
Both coefficients contain the factor $\lambda^2$. Substituting them into the integrated generator gives
\begin{equation}\label{eq:cumulative-dissipation}
\begin{aligned} D(T)&=\int_0^T\Gamma_2^{\rm can}(t)dt\succeq0,\\ \mathcal K_2(T)&=-i[H_2(T),\cdot]+\cD_{D(T)}.\end{aligned}
\end{equation}

Here the dissipator is defined as
\begin{equation}\label{eq:cumulative-dissipator}
 \cD_D(X)=\sum_{ab}D_{ab}\left(F_aXF_b-\tfrac12\{F_bF_a,X\}\right).
\end{equation}
Its size is controlled by the positive operator
\begin{equation}\label{eq:cumulative-radius}
 R_D=\sum_{ab}D_{ab}F_bF_a\succeq0,
 \, r_T=\norm{R_D}_\infty\le\Tr D(T).
\end{equation}
In particular, $\norm{\cD_D}_\diamond\le2r_T$. The same $r_T$ sets the normalizations $1+2r_T$ for the two-branch sampler and $e^{2r_T}$ for the exponential inverse of the cumulative model.

The integral of the rate matrix is positive. To establish this property, define $Y_a=\sum_\alpha\int_0^T a_{a\alpha}(t)B_\alpha(t)dt$. Then $D_{ab}=\lambda^2\Tr(\tau_BY_bY_a)$ is a positive Gram matrix. This cumulative GKSL structure also arises in dynamical coarse graining \cite{Schaller2009,Rivas2017}.

Assume that the canonical decoherence matrix
$\Gamma_2^{\rm can}(t)$ is integrable on every finite time interval.
At each time, we separate its positive and negative rates while
retaining the operator direction associated with each rate. In the
spectral decomposition of $\Gamma_2^{\rm can}(t)$, replacing each
eigenvalue $\gamma$ by $[\gamma]_+$ gives
$\Gamma_{2,+}^{\rm can}(t)$, and replacing it by $[-\gamma]_+$
gives $\Gamma_{2,-}^{\rm can}(t)$, where $[x]_+=\max(x,0)$.
This construction remains well defined when eigenvalues cross.
We then integrate both matrices in the fixed operator basis $\{F_a\}$,
\begin{equation}\label{eq:signed-general-dissipation}
 D(T)=D_+(T)-D_-(T),\,\;
 D_\pm(T)=\int_0^T\Gamma_{2,\pm}^{\rm can}(t)\,dt .
\end{equation}

One sees that a negative rate can offset an earlier positive contribution along the same direction. Both terminal corrections use the resulting net matrix $D(T)$ through \Cref{eq:cumulative-dissipation}.
Negative canonical rates of an exact, invertible differentiable evolution
witness failure of CP divisibility~\cite{Hall}. The quantities $c_j(Z)$
in \Cref{eq:price-path} show how recovery can affect the PEC cost
comparison in \Cref{eq:optimal-costs}. The connection is explicit for
scalar dephasing with $\dot\rho=\gamma(t)(Q\rho Q-\rho)$. For the pair
of states whose distinguishability is $e^{-\chi(t)}$, we have
\[
 c_j(Z)=\chi(t_j)-\chi(t_{j-1})
       =2\int_{t_{j-1}}^{t_j}\gamma(t)\,dt .
\]
An interval with $c_j(Z)<0$ restores distinguishability. Its contribution
reduces the end-of-circuit normalization, while independent sampling
retains the cost of the preceding loss. For the general second-order
dynamics considered here, the recovery seen by a particular pair depends
on how that pair evolves relative to the negative-rate directions.

\subsection{Implementing the correction at readout}
Fix an observable $O=O^\dagger$ with $\norm{O}_\infty\le1$.
Its ideal expectation is $\Tr(\rho_0 B)$, where $B=U_T^\dagger O U_T$.
Suppose a correction $\mathcal G_s$ has been designed for the
interaction-frame state $\widetilde\Phi_T(\rho_0)$. The device produces
$\Phi_T(\rho_0)=\mathcal U_T\widetilde\Phi_T(\rho_0)$, so its physical
output correction must satisfy
\begin{equation}\label{eq:branch}
 \overline{\mathcal G}_s\mathcal U_T=\mathcal U_T\mathcal G_s,
 \qquad
 \overline{\mathcal G}_s=\mathcal U_T\mathcal G_s\mathcal U_T^{-1}.
\end{equation}
This is a change of frame for compiling the correction. For example, if
$\mathcal G_s$ is conjugation by a Pauli $P$ and $U_T$ is Clifford, one
computes the Pauli $U_T P U_T^\dagger$ classically and applies it at
readout. There is no need to undo and repeat the ideal circuit. A
terminal channel acting only on the register can be applied to its reduced
state even when the bath remains correlated with the register, since the
channel commutes with the partial trace over the bath.

There are two ways to obtain a corrected expectation. One can append the
compiled channel and measure $O$, or measure its adjoint action on $O$
directly on the unmodified output. For a signed sum of terminal channels
with weights $w_s$, the two procedures give the same first moment,
represented by the   observable
$M=\sum_s w_s\overline{\mathcal G}_s^\dagger(O)$.
More generally, each implementation below constructs its own $M$ to meet
the same terminal objective,
\begin{equation}\label{eq:terminal}
 \sup_{\rho_0}\bigl|\Tr[M\Phi_T(\rho_0)]-\Tr[B\rho_0]\bigr|
 =\norm{\Phi_T^\dagger(M)-B}_\infty\le\epsilon_{\rm bias}.
\end{equation}

We consider two ways to implement the terminal correction. For a single
target observable, we compile the correction into the measurement. To
use the same correction for every target, we instead apply a sampled
channel before measuring the original observable. Both implementations
run the prescribed noisy circuit $\Phi_T$, as shown below.

\par\smallskip
\noindent\begin{minipage}{\columnwidth}
\centering
{\footnotesize
\setlength{\unitlength}{1pt}%
\begin{picture}(238,49)
\put(48,36){\makebox(34,10){Circuit}}
\put(88,36){\makebox(92,10){Output operation}}
\put(192,36){\makebox(46,10){Readout}}
\put(0,19){\makebox(50,14)[l]{Pauli readout}}
\put(0,1){\makebox(50,14)[l]{Channel pair}}
\multiput(52,19)(0,-18){2}{\framebox(27,14){$\Phi_T$}}
\multiput(79,26)(0,-18){2}{\vector(1,0){10}}
\put(89,19){\framebox(91,14){Rotate locally for $Q_j$}}
\put(89,1){\framebox(91,14){$\overline{\mathcal P}_T$ or
                                $\overline{\mathcal C}_T$}}
\multiput(180,26)(0,-18){2}{\vector(1,0){10}}
\put(190,19){\framebox(47,14){Read $Z$ bits}}
\put(190,1){\framebox(47,14){Measure $O$}}
\end{picture}}
\end{minipage}
\par\smallskip

In the first row, Algorithm~\ref{alg:observable-compiler} samples a
Pauli string $Q_j$ from the expansion of the corrected observable.
Local basis rotations and computational-basis readout give the outcome
of $Q_j$ by multiplying the measured signs on its support. In the
second row, \Cref{eq:two-branch-correction} specifies the probabilities
for sampling one of the two terminal channels. We then measure $O$.
In both cases, the sampled term determines the classical weight
applied to the measurement outcome.

\subsection{Correcting one observable by Pauli sampling}
For a single observable, we apply the adjoint of the formal inverse $\id-\mathcal K_2(T)$ to the target and transport the result to the output frame. 

In \Cref{eq:cumulative}, $\mathcal K_2(T)$ includes the factor
$\lambda^2$. We use $\rho_3(T)$ to denote a bound on the remaining error,
\begin{equation}\label{eq:remainder-certificate}
 \norm{\mathcal R_3(T)}_\diamond
 =\norm{\widetilde\Phi_T-\mathcal I-\mathcal K_2(T)}_\diamond\leq \rho_3(T).
\end{equation}
For a bounded joint interaction, put
$a_T=2|\lambda|\int_0^T\|V(t)\|_\infty\,dt$.
Iterating Duhamel's formula through second order gives
the choice $\rho_3(T)=a_T^3/3!$, since the exact joint propagator in the
remainder has diamond norm one. For unbounded interactions, a certificate can instead be obtained from bounds on higher bath correlations. 

We measure the size of the cumulative second-order error by
\begin{equation}\label{eq:kappaT}
 \kappa_T=\norm{\mathcal K_2(T)}_\diamond.
\end{equation}
The interaction-frame correction and its   observable are
\begin{equation}\label{eq:linear-observable-compiler}
 O_T^{(2)}=B-\mathcal K_2(T)^\dagger(B),\qquad
 \overline O_T^{(2)}=U_TO_T^{(2)}U_T^\dagger.
\end{equation}
The identity $\widetilde\Phi_T=\id+\mathcal K_2+\mathcal R_3$ gives
\begin{equation}\label{eq:linear-observable-bias}
 \|\widetilde\Phi_T^\dagger(O_T^{(2)})-B\|_\infty
 \le\kappa_T^2+(1+\kappa_T)\rho_3(T).
\end{equation}
This can be seen by composing $\widetilde\Phi_T^\dagger$ with $\id-\mathcal K_2^\dagger$, which cancels the order-$\lambda^2$ error. The residual consists of $-(\mathcal K_2^\dagger)^2 B$ and the Dyson remainder acting on $O_T^{(2)}$. The measurement correction thus uses the net effect of the bath on the target over the full circuit.

We now show how to compute the Pauli coefficients of the corrected
observable $O_T^{(2)}$ from the bath correlations and the ideal gate
sequence. Write each system coupling as $A_\alpha=g_\alpha P_\alpha$,
where $P_\alpha$ is a Pauli string and $g_\alpha$ is its coupling
strength.  Let the ideal circuit consist of instantaneous Clifford layers separated by noisy
intervals $I_l=[t_{l-1},t_l)$, $l=1,\dots,L$. On $I_l$ the toggled coupling
is $g_\alpha P_{\alpha,l}$, with the signed Pauli string
$P_{\alpha,l}=U_{t_{l-1}}^\dagger P_\alpha U_{t_{l-1}}$. Define the interval
integrals of the BCF,
\begin{equation}\label{eq:interval-bcf}
\begin{aligned}
 \mathcal G^{ll'}_{ab}&=\lambda^2\!\int_{I_l}\!dt\!\int_{I_{l'}}\!ds\,
 C_{ab}(t-s),\\
 \mathcal T^{l}_{ab}&=\lambda^2\!\int_{I_l}\!dt\!\int_{t_{l-1}}^{t}\!ds\,
 C_{ab}(t-s). 
\end{aligned}
\end{equation}
 Further, if the Pauli string $P_\alpha$ couples to the bath operator
$B_{b(\alpha)}$, the coefficient for the pair of strings $P_\alpha$
and $P_\beta$ on intervals $I_l$ and $I_{l'}$ is
\[
 w^{ll'}_{\alpha\beta}
 = g_\alpha g_\beta\,\mathcal G^{ll'}_{b(\alpha)b(\beta)} .
\]
The bath labels $b(\alpha)$ and $b(\beta)$ may coincide when the two
strings couple to the same bath operator.

Then
\Cref{eq:linear-observable-compiler} reads
\begin{align}
 O_T^{(2)}&=B-\sum_{l,l'}\sum_{\alpha\beta}
 w^{l'l}_{\beta\alpha}P_{\beta,l'}BP_{\alpha,l}
 +\Xi_T^\dagger B+B\,\Xi_T ,\label{eq:observable-pauli}
\end{align}
where 
\begin{align*}
    \Xi_T&=\sum_{l}\sum_{\alpha\beta}\Bigl[\sum_{l'<l}
 w^{ll'}_{\alpha\beta}P_{\alpha,l}P_{\beta,l'}
 +g_\alpha g_\beta\mathcal T^{l}_{b(\alpha)b(\beta)}
 P_{\alpha,l}P_{\beta,l}\Bigr].
\end{align*}
The Lamb shift is contained in $\Xi_T$,
$H_2(T)=(\Xi_T-\Xi_T^\dagger)/(2i)$. Every term is a Pauli string with a
complex coefficient, and Hermiticity makes the merged coefficients real.
Calibration requires the ordered complex correlations for each coupled pair of bath operators at the separations in \Cref{eq:interval-bcf}, including the time-ordered integral within an interval. For a quantum bath, Ramsey attenuation alone does not determine the antisymmetric response or all cross spectra. Correlation-sensitive spectroscopy can supply these additional inputs~\cite{VonLupke2020,PazSilva2017Noise}.
For a fitted model with $\|\widehat{\mathcal K}_2-\mathcal K_2\|_\diamond\leq\Delta_{\rm model}$, compile $B-\widehat{\mathcal K}_2^\dagger B$ and put
$\widehat\kappa_T=\|\widehat{\mathcal K}_2\|_\diamond$. For fits to $D(T)$ and $H_2(T)$, we may take $\Delta_{\rm model}=\Delta_{\rm cal}$ from \Cref{eq:cumulative-calibration}.
If the implemented measurement of $Q_j$ differs from its ideal response
by at most $\epsilon_j$, its contribution is bounded by
$\epsilon_{\rm ro}=\sum_j|a_j|\epsilon_j$.

\begin{algorithm}[tbp]
\caption{Corrected observable by Pauli sampling}
\label{alg:observable-compiler}
\textbf{Input:} a target $O$ whose interaction-frame operator
$B=U_T^\dagger OU_T$ is a Pauli string, the Clifford layers and noisy
intervals, the couplings $\{g_\alpha,P_\alpha,b(\alpha)\}_{\alpha=1}^m$, the
BCF with certificates $\Delta_{\rm model}$ and $\rho_3(T)$, and readout bounds $\epsilon_j$. Specify the statistical tolerance $\epsilon$ and failure probability $\delta$.\\
\textbf{Output:} an estimate $\widehat\mu$ of $\Tr(\rho_0B)$.
\begin{enumerate}
\item \emph{Transport.} For every interval $l$ and coupling $\alpha$,
compute the signed Pauli string $P_{\alpha,l}$.
\item \emph{Integrate.} Evaluate $\mathcal G^{ll'}$ and $\mathcal T^{l}$ of
\Cref{eq:interval-bcf}. For a stationary BCF and equal intervals, $\mathcal G^{ll'}$ depends only on $l-l'$, and $\mathcal T^l$ is the same on each interval.
\item \emph{Assemble.} Evaluate \Cref{eq:observable-pauli} as a list
of Pauli strings, merge equal strings, and conjugate them by $U_T$ to obtain
$\overline O_T^{(2)}=cI+\sum_ja_jQ_j$ with real $a_j$. Put
$K_{\rm meas}=\sum_j|a_j|$. If $K_{\rm meas}=0$, return $c$ without sampling.
\item \emph{Sample.} For each of
$N\ge2K_{\rm meas}^2\epsilon^{-2}\log(2/\delta)$ shots, draw $j$ with
probability $|a_j|/K_{\rm meas}$, run the unmodified noisy circuit, measure
$Q_j$ with outcome $y=\pm1$, and record
$c+K_{\rm meas}\operatorname{sgn}(a_j)\,y$.
\item \emph{Return} the average $\widehat\mu$ of the records.
\end{enumerate}
\textbf{Guarantee:} conditional on the model and readout certificates,
independent shots give, with probability at least $1-\delta$,
\[
 \begin{aligned}
 |\widehat\mu-\Tr(\rho_0B)|&\le\widehat\kappa_T^2+\epsilon_{\rm ro}+\epsilon
  +(1+\widehat\kappa_T)[\rho_3(T)+\Delta_{\rm model}].
 \end{aligned}
\]
Certificate failure probabilities add to $\delta$.\\
\textbf{Resources:} $O(m^2L^2)$ Pauli products before merging. Each shot uses at most $n$ single-qubit basis rotations and requires neither an ancilla nor an entangling readout gate.
\end{algorithm}
The Pauli implementation has shot overhead denoted by $K_{\rm meas}^2$.
For an exact model, let $s_T$ count the distinct nonidentity Pauli terms in the correction to the ideal output observable. Clifford conjugation takes the ideal Pauli target to a single Pauli. Orthogonality of the Pauli strings and Cauchy--Schwarz then give $K_{\rm meas}\le 1+\sqrt{s_T}\,\kappa_T$ (Appendix~\ref{app:pauli-sampling}). 

The compiled list of coefficients, sampled Pauli observables, and measurement bases can be represented using the circuit and observable primitives of QEM software such as Mitiq \cite{Mitiq}. Eddins et al.\ use Pauli propagation to absorb learned layerwise inverses into a measurement \cite{Eddins}. Here the coefficients carry the cross-interval bath integrals in \Cref{eq:interval-bcf}. Importantly, strings produced at different times can cancel before the sampling probabilities are assigned.

\subsection{Correcting all observables with two terminal quantum channels}
A Hermiticity-preserving, trace-preserving map can be represented as a
weighted difference of two CPTP maps~\cite{Jiang,Regula2021}.
Rossini et al.\ used this representation to implement a single-qubit
error-mitigation protocol~\cite{Rossini2023}. Here we construct the two terminal channels directly from the bath correlations accumulated
over the circuit. Their weights are $1+r_T$ and $-r_T$, where
$r_T=\norm{R_D}_\infty$ from \Cref{eq:cumulative-radius} is computed after contributions from
decoherence and recovery have been combined in $D(T)$. The resulting
normalization $1+2r_T$ applies to every target observable, with the
fixed-time bias bounded above by \Cref{eq:cumulative-bias}.

We first
diagonalize the integrated matrix,
\[ D(T)=\sum_\mu d_\mu \ketbra{v_\mu} \]
 and subsequently define
\begin{equation}\label{eq:cumulative-jumps}
 L_\mu=\sqrt{d_\mu}\sum_a  \braket{a}{v_\mu} F_a,
 \,\, \mathcal J_D(X)=\sum_\mu L_\mu XL_\mu^\dagger.
\end{equation}
Then $R_D=\sum_\mu L_\mu^\dagger L_\mu$ and $r_T=\norm{R_D}_\infty$ agree with \Cref{eq:cumulative-radius}.

For $r_T>0$, we define
\begin{align}
 W_T&=\sqrt{r_TI-R_D},\nonumber\\
 \mathcal C_T(X)&=\frac{\mathcal J_D(X)+W_TXW_T}{r_T},\label{eq:completion-channel}\\
 \mathcal L_{0,T}&=+i[H_2(T),\cdot]+\cD_{W_T},\nonumber\\
 \cD_{W_T}(X)&=W_TXW_T-\tfrac12\{W_T^2,X\}.
 \label{eq:positive-drift}
\end{align}
The channel $\mathcal C_T$ is clearly CPTP. Further, $\mathcal L_{0,T}$ is a GKSL generator, and
\begin{equation}\label{eq:inverse-completion}
 -\mathcal K_2(T)=\mathcal L_{0,T}+r_T(\id-\mathcal C_T).
\end{equation}
For $r_T=0$, set $W_T=0$ and $\mathcal L_{0,T}=+i[H_2(T),\cdot]$. The deterministic coherent correction then suffices at this order.

We notice that $L_\mu/\sqrt{r_T}$ and $W_T/\sqrt{r_T}$ are Kraus operators of
$\mathcal C_T$, and
$e^{u\mathcal L_{0,T}}$ is physical for $u\ge0$. We apply the stochastic-PEC
completion \cite{Sun2021,Donvil2025} to the integrated generator:

\begin{equation}\label{eq:two-branch-correction}
\begin{aligned} \mathcal P_T&=e^{\mathcal L_{0,T}/(1+r_T)}, \;  \mathcal R_T^{(2)}=(1+r_T)\mathcal P_T-r_T\mathcal C_T.\end{aligned}
\end{equation}
Expanding the physical channel $\mathcal P_T$ in $u$ gives
\begin{equation}\label{eq:explicit-leading-cancellation}
\begin{aligned}\mathcal R_T^{(2)}&=\id+\mathcal L_{0,T}+r_T(\id-\mathcal C_T)+O(\lambda^4)
\\ &=\id-\mathcal K_2(T)+O(\lambda^4).\end{aligned}
\end{equation}
The sampled channels in \Cref{eq:two-branch-correction} therefore reproduce the required inverse through second order. In this use of PEC \cite{Temme}, the positive and negative weights are assigned after the bath correlations have been integrated over the complete circuit.

\Cref{eq:two-branch-correction} suggests that we 
choose $\mathcal P_T$ with probability $(1+r_T)/(1+2r_T)$ and sign $+1$, or $\mathcal C_T$ with probability $r_T/(1+2r_T)$ and sign $-1$. Append the chosen channel in the output frame of \Cref{eq:branch}, measure $O$, and multiply the outcome by its sign and $1+2r_T$. Average these scores over independent shots. When $r_T=0$ only the deterministic coherent correction remains. For $r_T>0$, the Kraus representation of $\mathcal C_T$ uses at most $\operatorname{rank}D(T)+1$ operators and admits a dilation with $\lceil\log_2(\operatorname{rank}D(T)+1)\rceil$ fresh ancillas. This count applies to $\mathcal C_T$. Implementing $\mathcal P_T$ also requires synthesizing the drift in \Cref{eq:positive-drift} in the output frame. Its depth depends on $H_2(T)$, $W_T$, and the available gates. Since $W_T$ is Hermitian, the drift branch has a simple implementation to the required order. Write $b=(1+r_T)^{-1}$, $\mathcal H(X)=i[H_2(T),X]$, and draw $\theta\sim\mathcal N(0,b)$. Applying $e^{-i\theta W_T}$ followed by $e^{ibH_2(T)}$ realizes $e^{b\mathcal H}e^{b\cD_{W_T}}$ on average~\cite{Salgado2002}. The signed weights are unchanged. The added bias is at most $b\norm{[\mathcal H,\cD_{W_T}]}_\diamond/2=O(\lambda^4)$ at fixed duration (Appendix~\ref{app:pauli-sampling}).

We write $K_{\rm meas}(B)$ for the normalization of the Pauli
measurement sampler for target $B$, and $K_{\rm lin}$ for the
target-independent normalization of the two-channel sampler. 

\begin{theorem}[Cost and bias of the two terminal corrections]\label{thm:cumulative-cost}
Assume the centered, factorized-bath expansion in \Cref{eq:cumulative}
with exact model inputs and independent shots. With the remainder bound
and cumulative size from \Cref{eq:remainder-certificate,eq:kappaT}, put
\[
 E_T=\kappa_T^2+(1+\kappa_T)\rho_3(T).
\]

\emph{(i) Pauli readout.} For the Clifford circuit, Pauli couplings and
target of Algorithm~\ref{alg:observable-compiler}, write its corrected
  observable as $cI+\sum_j a_j Q_j$. With ideal Pauli readout
the normalization is $K_{\rm meas}(B)=\sum_j|a_j|$ and the uniform bias for
this target is at most $E_T$.

\emph{(ii) Terminal channels.} Suppose the two channels in
\Cref{eq:two-branch-correction} can be implemented as specified, and
write $\ell_T=\norm{\mathcal L_{0,T}}_\diamond$ for the drift in \Cref{eq:positive-drift}. Their signed sampler has
\begin{equation}\label{eq:cumulative-cost}
 K_{\rm lin}=1+2r_T.
\end{equation}
The same channels correct every target of norm at most one with bias
bounded by
\begin{equation}\label{eq:cumulative-bias}
 \norm{\mathcal R_T^{(2)}\widetilde\Phi_T-\id}_\diamond
 \le E_T+\frac{\ell_T^2}{2(1+r_T)}.
\end{equation}

For either sampler, $N\ge 2K^2\epsilon^{-2}\log(2/\delta)$ shots give
statistical error at most $\epsilon$ with probability at least $1-\delta$,
using the normalization of that sampler. At fixed $T$, if $\rho_3(T)=O(|\lambda|^3)$, both ideal systematic biases are $O(|\lambda|^3)$. If the cubic reduced coefficient vanishes and the fourth-order remainder is $O(\lambda^4)$, both biases are $O(\lambda^4)$.
\end{theorem}

The two constructions have different costs when the target observable
is sensitive to noise on only part of the register. To see this,
consider $n$ qubits subject to independent dephasing. 
 Suppose $H_2=0$ and
$\mathcal K_2=b(T)\sum_{j=1}^n(\Ad_{Z_j}-\id)$, where $b(T)\ge0$ is
the integrated strength per qubit. Take a nonidentity Pauli target $B$ with $X$ or
$Y$ on $w$ sites and a Clifford output frame. Only those $w$ terms
contribute to $\mathcal K_2^\dagger(B)=-2wb(T)B$, whereas the
channel correction has $R_D=nb(T)I$. The two shot overheads are
\begin{equation}\label{eq:target-size-overhead}
 K_{\rm meas}(B)^2=[1+2w b(T)]^2,\,\,
 K_{\rm lin}^2=[1+2n b(T)]^2 .
\end{equation}
Thus the Pauli-readout overhead depends only on the number of sites
where dephasing attenuates the target. Even a Pauli string supported
on the entire register, such as $X_1Z_2\cdots Z_n$, has $w=1$.
The two-channel sampler pays for dephasing on all $n$ sites because
its correction must work for every observable.

Finally, the size of $r_T$ is determined by the bath correlations accumulated
under the prescribed controls. In the next subsection,
\Cref{eq:bounded-spectrum-cost} bounds its growth with circuit duration,
and \Cref{prop:shifted-moments} identifies spectral conditions under
which it remains bounded uniformly in $T$.

\subsection{Sampling overhead from the bath spectrum}

As in the commuting Gaussian case, the sampling overhead depends on how
the bath spectrum overlaps the frequencies selected by the ideal
evolution. For the general couplings considered here, each coupling
operator has its own frequency response. We use these responses to
identify two regimes. A bounded spectral density gives a polynomial
shot bound for the two-branch correction, while sufficiently weak
spectral weight near the coupled transition frequencies gives a bound
independent of the circuit duration.

Write the ordered bath correlation function in \Cref{eq:bcf} as
$C(\tau)=\int_{\mathbb R}e^{-i\omega\tau}J(\omega)\,d\omega$,
where $J(\omega)\succeq0$. For each coupling operator, define the
Fourier transform of its interaction-frame evolution over the circuit,
\begin{equation}\label{eq:operator-filter}
 \widehat A_{\alpha,T}(\omega)
 =\int_0^T e^{i\omega t}\widetilde A_\alpha(t)\,dt .
\end{equation}
Contributions from different times are added in
$\widehat A_{\alpha,T}(\omega)$ before their magnitude enters the cost.
To see this, insert the spectral representation of the BCF into
\Cref{eq:cumulative-gram,eq:cumulative-radius}. The result is
\begin{equation}\label{eq:operator-filter-radius}
 \begin{aligned}
 S_T(\omega)
 &=\sum_{\alpha,\beta}[J(\omega)]_{\alpha\beta}\,
   \widehat A_{\alpha,T}(\omega)^\dagger
   \widehat A_{\beta,T}(\omega),\\
 R_D&=\lambda^2\int_{\mathbb R}S_T(\omega)\,d\omega .
 \end{aligned}
\end{equation}
Since $J(\omega)$ is positive semidefinite, so is $S_T(\omega)$.
Taking the operator norm gives the spectral bound
\begin{equation}\label{eq:general-filter-budget}
 r_T=\norm{R_D}_\infty
 \le b_T
 :=\lambda^2\int_{\mathbb R}\norm{S_T(\omega)}_\infty\,d\omega
 \le \Tr D(T).
\end{equation}
Inserting this estimate into \Cref{eq:cumulative-cost} gives the two-branch shot bound $(1+2b_T)^2$ while retaining cancellations in the coupling history.

\noindent\textbf{Polynomial overhead for a bounded spectrum.}
Suppose $J_{\max}:=\operatorname*{ess\,sup}_{\omega}
\norm{J(\omega)}_\infty<\infty$. For any unit vector $\psi$, we have
\[
 \langle\psi|S_T(\omega)|\psi\rangle
 \le J_{\max}\sum_\alpha
       \norm{\widehat A_{\alpha,T}(\omega)\psi}^{2}.
\]
Plancherel's identity converts each frequency integral on the right
into $2\pi\int_0^T\norm{\widetilde A_\alpha(t)\psi}^2dt$.
Unitary evolution preserves $\norm{A_\alpha}_\infty$, so taking the
supremum over $\psi$ gives the operator-norm bound
\begin{equation}\label{eq:bounded-spectrum-cost}
 r_T\le 2\pi\lambda^2 J_{\max}T
             \sum_\alpha\norm{A_\alpha}_\infty^2 .
\end{equation}
For an $n$-qubit register with $O(n)$ coupling operators of uniformly
bounded norm and $J_{\max}$ independent of $n$, the two-branch shot
overhead is at most $[1+O(\lambda^2 nT)]^2$. Thus, at fixed register
size, coupling strength, and statistical precision, the number of shots
needed to estimate the corrected expectation grows at most quadratically
with the circuit duration. To estimate the ideal expectation to a
prescribed total accuracy, one must also control the bias of the
second-order correction at that duration, as quantified in
\Cref{eq:cumulative-bias}.

\noindent{\bf Bounded overhead.}
For a time-independent ideal Hamiltonian, the relevant frequencies
can be read off directly from the coupling operators. Write
$\widetilde A_\alpha(t)=\sum_{\nu\in V_\alpha}
e^{i\nu t}A_{\alpha\nu}$, where $V_\alpha$ contains the Bohr
frequencies at which $A_\alpha$ couples energy eigenspaces. 
Using the finite-time Fourier integral $F_T$ from
\Cref{eq:spectral-covariance}, we obtain
\[
 \widehat A_{\alpha,T}(\omega)
 =\sum_{\nu\in V_\alpha} F_T(\omega+\nu)A_{\alpha\nu}.
\]
A bath frequency $\omega$ is resonant with the system component
$A_{\alpha\nu}$ when $\omega+\nu=0$. At resonance the phase
$e^{i(\omega+\nu)t}$ is constant, so $F_T(\omega+\nu)=T$.
For nonzero detuning, the oscillations instead give the uniform bound
\[
 |F_T(\omega+\nu)|
 =\left|\frac{e^{i(\omega+\nu)T}-1}{i(\omega+\nu)}\right|
 \le \frac{2}{|\omega+\nu|}.
\]
 The following result uses these
bounds to identify spectra for which extending the evolution leaves
the sampling overhead bounded.

\begin{proposition}[Bounded sampling overhead away from resonance]
\label[proposition]{prop:shifted-moments}
For a time-independent ideal Hamiltonian on a finite-dimensional register, define the inverse-square moment about each coupled resonance $-\nu$ by
\begin{equation}\label{eq:shifted-moment}
 B_\nu=\int_{\mathbb R}
       \frac{\norm{J(\omega)}_\infty}{|\omega+\nu|^2}\,d\omega .
\end{equation}
If all these moments are finite, put
\begin{equation}\label{eq:shifted-budget}
 B_{\mathrm{shift}}
 =4\sum_\alpha\left(
    \sum_{\nu\in V_\alpha}\norm{A_{\alpha\nu}}_\infty\sqrt{B_\nu}
                \right)^2 .
\end{equation}
Then, for every $T\ge0$,
\begin{equation}\label{eq:shifted-cost}
 \begin{aligned}
 r_T&\le\lambda^2 B_{\mathrm{shift}}, \quad 
 K_{\mathrm{lin}}^2
 \le(1+2\lambda^2 B_{\mathrm{shift}})^2 .
 \end{aligned}
\end{equation}
The same spectral bound applies if we use \Cref{eq:poisson-inverse}
to invert the cumulative model $e^{\mathcal K_2(T)}$ once at readout.
The shot overhead of this representation is
$K_{\mathrm{cum}}^2=e^{4r_T}\le e^{4\lambda^2 B_{\mathrm{shift}}}$.
\end{proposition}

\Cref{eq:shifted-cost} bounds the sampling overhead uniformly in duration. To certify a prescribed total error at a chosen $T$, the systematic contribution in \Cref{eq:cumulative-bias} must also fit the error budget. For a bath of finite spectral weight, a gap around every coupled
resonance $-\nu$ suffices. For discrete bath modes, the moments in
\Cref{eq:shifted-moment} are sums over the spectral lines. 
For the fixed commuting couplings of \Cref{sec:spectral}, only
$\nu=0$ contributes. In that case \Cref{thm:spectral} gives the
sharper result that a finite inverse-square moment in each direction
selected by the observable is both necessary and sufficient for
bounded exact overhead.

\subsection{A brief comparison of sampling costs}
We use independent cancellation as a benchmark for the cost of the
terminal correction. With the positive and negative rate matrices from
\Cref{eq:signed-general-dissipation}, write
$R_\pm(t)=\sum_{ab}[\Gamma_{2,\pm}^{\rm can}(t)]_{ab}F_bF_a$ and
$r_+(t)=\norm{R_+(t)}_\infty$. Applying the same channel completion
independently to each inverse increment gives the coefficient benchmark
$K_{\mathrm{layer},2}=\exp(2\int_0^T r_+(t)\,dt)$. The terminal
construction combines the contributions first. Since
$R_D=\int_0^T(R_+(t)-R_-(t))\,dt\succeq0$, we have
$r_T=\norm{R_D}_\infty\leq\int_0^T r_+(t)\,dt$, and hence
\begin{equation}\label{eq:cumulative-versus-local}
\begin{aligned}
 K_{\rm lin}^2=&(1+2r_T)^2
 \leq K_{\rm cum}^2=e^{4r_T}\\
 &\leq K_{\mathrm{layer},2}^2
   =\exp\!\left(4\int_0^T r_+(t)\,dt\right).
\end{aligned}
\end{equation}
\Cref{eq:cumulative-versus-local} compares the sampling
normalizations achieved by the two constructions. For noncommuting
generators, the cumulative exponential and the time-ordered TCL2
propagator agree through second order and can differ at fourth order
at fixed $T$.

Wang and Li obtain a related shot bound from an absolute envelope of
the bath correlations. Taking the fine-grid limit of the one-step and product bounds in Eqs.~(51) and (52) of their arXiv version~\cite{WangLi} and squaring the normalization gives, in their coupling convention with gate-library constant $Q$,
\begin{equation}\label{eq:bcf-envelope-comparator}
\begin{aligned}
    K_{\rm WL}^2 & \leq \exp\!\left[4\lambda^2 Q\,\mathcal A_C(T)\right],\\
 \mathcal A_C(T)&=\int_0^T(T-\tau)
       \sum_{\alpha\beta}|C_{\alpha\beta}(\tau)|\,d\tau .
\end{aligned}
\end{equation}
The absolute value counts each delay separately. In
\Cref{eq:cumulative-gram}, we integrate the bath correlations against
the prescribed control history before taking the norm. Contributions
from different times can then cancel, yielding a smaller cost bound
even when $\mathcal A_C(T)$ grows with the circuit duration.
Appendix~\ref{app:many-proof} gives two analytical separations for noncommuting couplings, one at fixed duration and one at repeated full revivals.

\section{Numerical results}\label{sec:numerics}
The first two tests use the same polarized bath qubit to examine the two corrections of \Cref{thm:cumulative-cost}. In the first, we follow the normalization and the observable bias of the two terminal channels over many bath oscillations. In the second, we use the Pauli measurement compiler on a four-qubit gate circuit and compare bias and sampling error at a fixed shot budget. The third test applies the exact scalar-dephasing comparison of \Cref{eq:compression-costs} to single-qubit storage with and without a refocusing pulse. Appendix~\ref{app:numerical-methods} gives the retained-bath derivation and the protocols used in Tables~\ref{tab:pauli-main} and~\ref{tab:background-echo}.

\subsection{Cumulative cancellation in a retained bath}
\label{sec:numerics-explicit}
\begin{figure*}[t]
 \centering\includegraphics[width=\linewidth]{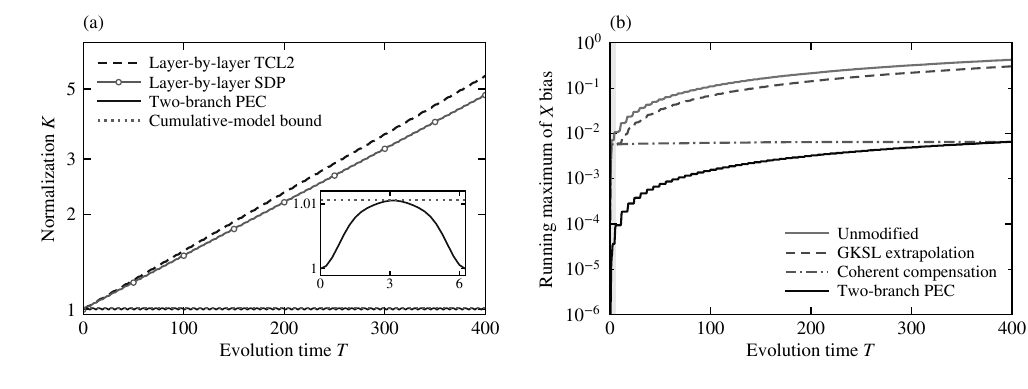}
 \caption{Retained polarized bath at $\lambda=0.08$. (a) Sampling normalization $K$ for two-branch PEC, the layer-by-layer TCL2 benchmark (dashed), and the optimal independent exact inverses at step boundaries $t_j=j/2$ (open circles). The dotted line bounds the cumulative-model normalization. The inset resolves the first oscillation. Shot overhead is $K^2$. (b) Largest systematic $X$ bias observed up to each time $T$ for the unmodified circuit, coherent compensation alone, two-branch PEC, and the inverse of a time-independent GKSL fit (The shaded interval contains the fit time). Each curve shows the running maximum of $b_X$  from \Cref{bX} over evaluation times spaced by $0.25$ with $\pi$ and $2\pi$ also included.}
 \label{fig:explicit-compiler}
\end{figure*}

Consider a data qubit with ideal Hamiltonian $Z_S/2$ coupled to a bath qubit that is retained throughout each circuit execution. The joint Hamiltonian is $H_0+\lambda V$, with
\begin{subequations}\label{eq:explicit-bath}
\begin{equation}
 \begin{aligned}
 H_0&=\tfrac12 Z_S+Z_B,\qquad \tau_B=\tfrac12(I+p_B Z_B),\\
 V&=\tfrac12 X_S X_B+0.35 Z_S Y_B,\qquad p_B=0.6 .
 \end{aligned}
\end{equation}
In the bath basis $(B_x,B_y)=(X_B,Y_B)$, put $c=\cos(2\tau)$ and $s=\sin(2\tau)$. Evolving $X_B,Y_B$ under $H_B=Z_B$ and taking the ordered expectation in \Cref{eq:bcf} gives the exact BCF
\begin{equation}\label{eq:retained-spin-bcf}
 C(\tau)=
 \begin{pmatrix}
 c+ip_B s & -s+ip_B c\\
 s-ip_B c & c+ip_B s
 \end{pmatrix}.
\end{equation}
\end{subequations}
The coupling strengths are included in the system operators $X_S/2$ and $0.35Z_S$. This non-Gaussian bath produces a nonunital reduced channel, and the system couplings do not commute. We integrate \Cref{eq:retained-spin-bcf} with the ideal control history, i.e., $U_t=e^{-itZ_S/2}$,  to obtain $H_2(T)$ and $D(T)$. These specify the two terminal channels $\mathcal P_T,\mathcal C_T$ and their signed weights $1+r_T,-r_T$ in \Cref{eq:two-branch-correction}. To test this second-order construction, we compute the exact reduced channel separately by propagating the full Hamiltonian and tracing out the bath.

Figure~\ref{fig:explicit-compiler}(a) compares sampling normalizations. Applying the same completion to each instantaneous inverse gives the coefficient benchmark
\begin{equation}\label{eq:numerical-local-benchmark}
 \begin{aligned}
 K_{\rm layer,2}&=\exp\!\left(2\int_0^T r_+(t)\,dt\right),\\
 r_+(t)&=\left\|\sum_{ab}[\Gamma_{2,+}^{\rm can}(t)]_{ab}F_bF_a\right\|_\infty .
 \end{aligned}
\end{equation}
The BCF has spectral lines at $\pm2$, separated from the coupled system frequencies $0,\pm1$. As predicted by Proposition~\ref{prop:shifted-moments}, the cumulative dissipator stays bounded. In this model it is periodic, and Appendix~\ref{app:retained-time} gives
\begin{equation}\label{eq:retained-uniform-cost}
 \begin{aligned}
 2r_T&\le\tfrac{74}{45}\lambda^2,\\
 K_{\rm lin}&\le1.010525,\quad K_{\rm cum}\le1.010581
 \quad(\lambda=0.08).
 \end{aligned}
\end{equation}
The positive-rate contribution to the layer-by-layer benchmark recurs each period, giving $\log K_{\rm layer,2}=0.664064\lambda^2 T+O(\lambda^2)$ and $K_{\rm layer,2}=5.50577$ at $T=400$. We also computed the optimal full-channel normalizations from the exact reduced channels at times $t_j=j/2$ using the semidefinite programs for \Cref{eq:optimal-costs}. At these times through $T=400$, $K_{\rm end}$ stays below $1.013$, while $K_{\rm layer}(400)=4.77879$.

Figure~\ref{fig:explicit-compiler}(b) tests the accuracy of the corrected $B=X$ expectation. For a terminal correction $\mathcal R_T$ in the interaction frame, the systematic bias maximized over input states is
\begin{equation}\label{bX}
    b_X(T;\mathcal R_T)=
 \bigl\|(\mathcal R_T\circ\widetilde\Phi_T)^\dagger(X)-X\bigr\|_\infty ,
\end{equation}
where $\widetilde\Phi_T$ is obtained from exact joint propagation. Taking $\mathcal R_T=\mathcal I$ gives the raw bias, and taking $\mathcal R_T=\mathcal R_T^{(2)}$ gives the bias remaining after two-branch PEC. At $T=6$, a coupling sweep gives fitted powers of $1.99814$ and $3.99917$ in $|\lambda|$ for these two biases. Conjugation by $Z_B$ reverses the sign of the coupling while preserving $H_0$ and $\tau_B$, so the odd reduced orders vanish, as required for the fourth-order conclusion of \Cref{thm:cumulative-cost}.

At $\lambda=0.08$, two-branch PEC reduces the worst-case $X$ bias
from $0.0530$ in the unmitigated circuit to $7.58\times10^{-4}$
at $T=50$. At $T=200$ the bias falls from $0.2124$ to
$3.22\times10^{-3}$, and at $T=400$ it falls from $0.4222$ to
$2.98\times10^{-3}$.  At $T=50$ the generic quartic remainder bound is $a_T^4/24=89.09$. 

There are two further comparisons in panel (b). The coherent-compensation curve in Fig.~\ref{fig:explicit-compiler}(b)
shows the effect of undoing only the rotation generated by $H_2(T)$.
This unitary correction has normalization $K=1$ and accounts for much
of the reduction in $X$ bias in this example. At $T=50$ and $T=200$,
it gives slightly smaller bias than two-branch PEC. Including the
dissipative correction gives the smaller pointwise bias at $91.3\%$
of the sampled times in $0<T\le200$. The other comparison fits a time-independent GKSL generator to exact reduced channels for $0<T\le6$ and applies its inverse noise model at later times. This measures how well a Markovian fit extrapolates beyond its calibration interval.

\subsection{Pauli readout in a four-qubit gate circuit}
\label{sec:layered-test}

We test Algorithm~\ref{alg:observable-compiler} on a four-qubit
Clifford circuit and ask how much accuracy is gained by correcting
the mixing of the target with other Pauli observables. We retain the
bath Hamiltonian, polarization, and ordered BCF of
\Cref{eq:retained-spin-bcf}. During each interval of duration $1/2$,
the four data qubits interact with the same bath qubit through
\[
 \begin{aligned}
 H&=I_S\otimes Z_B+\lambda(A_x\otimes X_B+A_y\otimes Y_B),\\
 A_x&=0.5Z_0+0.45Z_1+0.35Z_2+0.3Z_3,\\
 A_y&=0.25X_1+0.2X_3,\qquad \lambda=0.12 .
 \end{aligned}
\]
Each interval is followed by one of the two alternating Clifford layers
in \Cref{eq:layered-gates}, each containing single-qubit gates and two
CNOTs. The bath qubit is retained throughout the circuit. The Clifford
layers determine which system Pauli operators multiply the interval
BCF integrals in \Cref{eq:interval-bcf}. From these contributions, the
compiler constructs a corrected observable that can be sampled using
local Pauli measurements.

We choose the interaction-frame target $B=X_0$ and examine the circuit
at two consecutive depths. After 16 layers the ideal unitary returns
to the identity up to a phase. The 15-layer prefix implements a
nonidentity unitary but also preserves $X_0$. The same laboratory
observable therefore tests the compiler both before and at the ideal
circuit recurrence.

\begin{table}[hpb]
\centering\small
\begin{tabular}{@{}lrrr@{}}
\toprule
Depth & Raw & Scalar & Pauli compiler\\
\midrule
\multicolumn{4}{@{}l}{\emph{Worst-case absolute bias, exact model}}\\
15 layers & $0.02729$ & $0.02652$ & $0.000707$\\
16 layers & $0.03236$ & $0.03164$ & $0.000956$\\[2pt]
\multicolumn{4}{@{}l}{\emph{Shot overhead $K^2$, exact model}}\\
15 layers & $1$ & $1.0108$ & $1.2822$\\
16 layers & $1$ & $1.0142$ & $1.3488$\\[2pt]
\multicolumn{4}{@{}l}{\emph{Fixed-input MSE in units of $10^{-6}$}}\\
15 layers & $28.675$ & $7.3711$ & $9.5261$\\
16 layers & $33.822$ & $6.5751$ & $10.243$\\
\bottomrule
\end{tabular}
\caption{Readout of $X_0$ for the untwirled four-qubit circuit with
ideal measurement operations. The first two comparisons use exact
model inputs. The bias is maximized over input states, and the shot
overheads are $d^{-2}$ for scalar rescaling and $K_{\rm meas}^2$ for
the Pauli compiler. The final comparison gives the mean-square error
on $[R_y(\pi/4)\ket{0}]^{\otimes4}$ at a total budget of $10^5$ shots.
Raw readout uses all shots for production. Each corrected readout
uses $10^4$ calibration shots and $9\times10^4$ production shots.}
\label{tab:pauli-main}
\end{table}

As a baseline, we use \emph{scalar rescaling}. Let $d$ be the
coefficient of $X_0$ in $\widetilde\Phi_T^\dagger(X_0)$. Dividing the
raw $X_0$ measurement average by $d$ restores the coefficient of
$X_0$ to one. Other Pauli components can still contribute to the
measured mean, depending on the input state. The compiler corrects
this mixing through second order by sampling one Pauli term from the
corrected observable on each shot. When calibration is included, both
methods estimate $d$ from the same measurements on the prescribed
circuit. Scalar rescaling divides by $\widehat d$, while the compiler
uses $\widehat d$ to fit the coupling strength and construct the Pauli
coefficients, as described in Appendix~\ref{app:observable-practice}.

Table~\ref{tab:pauli-main} shows that compiling the other Pauli
components reduces the worst-case bias at both depths. At 16 layers,
the bias falls from $0.03164$ for scalar rescaling to $0.000956$ for
the compiler, a factor of $33.1$. The independent bound from
\Cref{eq:finite-bath-validation} at $p=16$ gives
$9.606\times10^{-4}$, close to the bias $9.559\times10^{-4}$
obtained by exact joint propagation. Although the compiled observable
contains 223 nonidentity Pauli terms, their coefficient norm is
$K_{\rm meas}=1.16140$, giving a shot overhead of $1.3488$.
On the specified product input, both corrections reduce the MSE
relative to raw readout. At the same total shot budget, scalar
rescaling gives the smaller MSE at both depths. The compiler's
advantage in this example is its accuracy across input states.

The same 16-layer circuit also tests the two-channel construction in
\Cref{eq:two-branch-correction}. With exact model inputs, compensating
only the coherent shift leaves a worst-case bias of $0.01949$.
Including the dissipative correction reduces this bias to $0.001023$,
close to that of the Pauli compiler. The two-channel sampler has
normalization $K_{\rm lin}=1.12420$ and applies one sampled channel at
readout. The operator-filter bound in \Cref{eq:general-filter-budget}
gives $b_T=0.06761$ and hence $K_{\rm lin}\le 1.13522$.

Finally, we repeat the 16-layer test with independent full Pauli
frames to compare two ways of learning the attenuation of $X_0$.
Calibrating the attenuation directly on the complete circuit gives
MSE $8.9320\times10^{-6}$. Fitting six Pauli rates on isolated
twirled intervals and propagating them through the circuit gives
$7.9433\times10^{-5}$. Each method uses 20000 calibration shots and
80000 production shots. With exact calibration, the layer model
predicts an attenuation of $0.966061$, close to the actual value
$0.966056$. The larger finite-shot MSE therefore comes chiefly from
estimating six rates and reusing them across the circuit.

\subsection{End-of-circuit mitigation after refocusing}
\label{sec:realistic-background}\label{sec:background-echo}
We now consider a single qubit storing a transverse $X$ signal in the Ramsey rotating frame \cite{Oda,OdaData,Burnett2019}. We compare two ideal control sequences of the same duration $T$. In \emph{free evolution} the qubit is idle, with $H_{\rm id}=0$ and $U_{\rm id}(T)=I$. In a \emph{midpoint echo} an instantaneous $X$ pulse at $T/2$ reverses the sign of the longitudinal noise coupling for the second half. The net ideal operation is $X$ up to phase and leaves the target $X$ invariant. For each sequence separately, we compare independent inversion over successive time steps with inversion at readout.

\begin{figure*}[tbp]
\centering\includegraphics[width=\textwidth]{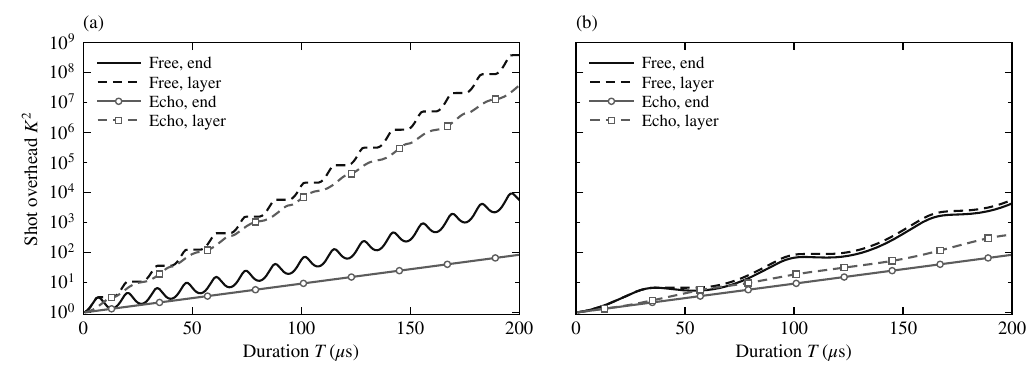}
\caption{Shot overhead for single-qubit storage with the faster fluctuator assignment (a) $v=0.23224$ and the slower assignment (b) $v=0.047936\,\mathrm{rad}/\mu\mathrm{s}$. Free evolution has no intermediate pulse. Echo applies one ideal $X$ pulse at $T/2$. Within each control sequence, solid and dashed curves give terminal and layer-by-layer inversion, respectively. Open circles and open squares identify the two echo curves. Both panels use the same white floor, quasistatic $1/f$ approximation, and nominal $0.5\us$ time steps, with the pulse time included as a step boundary.}
\label{fig:background-echo}
\end{figure*}

We model the longitudinal noise by combining a quasistatic fluctuator with values $\pm v$ and probabilities $(1\pm p)/2$, Gaussian $1/f$ frequency noise, and a white dephasing floor of rate $r$. The three contributions are independent. Writing $h$ for the multiplier of $\langle X\rangle+i\langle Y\rangle$ in the control frame, free evolution gives
\begin{equation}\label{eq:background-coherence}
 h(T)=e^{-rT-\chi_{1/f}(T)}[\cos(vT)+ip\sin(vT)] .
\end{equation}
The scalar comparison of \Cref{eq:compression-costs} applies directly to the attenuation $|h|$ after the coherence phase is absorbed into the readout basis. The fluctuator distribution gives the bracketed factor exactly. This synthetic study uses noise scales from separate experiments~\cite{Burnett2019,Oda,OdaData}. We choose $p=0.6$ and set the white floor $r=0.011097\,\mu\mathrm{s}^{-1}$ equal to the fitted Ramsey envelope decay as a model assumption. Both $v=0.23224$ and $0.047936\,\mathrm{rad}/\mu\mathrm{s}$ are compatible with the published $X$-only Ramsey fit. Appendix~\ref{app:realistic-background} gives the frequency band, parameter sources, and fit.

In free evolution, a static detuning accumulates phase for the full time $T$. Under echo its signed elapsed time is $a_T(t)=t$ before the pulse and $a_T(t)=T-t$ afterward. In the quasistatic approximation the echo exponent at intermediate time $t$ is $\chi_{\mathrm e,T}(t)=rt+f(a_T(t))$, with
\[
 f(a)=\tfrac12\sigma_\omega^2a^2-\tfrac12\log[\cos^2(va)+p^2\sin^2(va)] .
\]
At readout $a_T(T)=0$, so only the white dephasing floor survives.  To connect this recovery to the sampling cost, consider two opposite
equatorial input states. Their trace distance at time $t$ is the
coherence magnitude $|h_{\mathrm e,T}(t)|=e^{-\chi_{\mathrm e,T}(t)}$.
The scalar-dephasing identity in \Cref{eq:compression-costs} therefore
applies directly: terminal inversion pays for the attenuation at $T$,
while independent inversion accumulates the positive increments of
$\chi_{\mathrm e,T}$ along the echo sequence. Thus 
\begin{equation}\label{eq:echo-local-cost}
 \begin{aligned}
 K_{\mathrm e,end}^2&=e^{2rT},\\
 \Delta_j\chi_{\mathrm e,T}
 &=\chi_{\mathrm e,T}(t_j)-\chi_{\mathrm e,T}(t_{j-1}),\\
 K_{\mathrm e,layer}^2&=\exp\!\left(2\sum_j[\Delta_j\chi_{\mathrm e,T}]_+\right).
 \end{aligned}
\end{equation}
Both frequency assignments therefore have the same terminal echo overhead in Fig.~\ref{fig:background-echo}. Using the full finite-band $1/f$ filter changes the final exponent by less than $2.6\times10^{-11}$ through $200\us$.

\begin{table*}[tbp]
\centering\small
\caption{Matched comparisons at $T=100\us$ with $0.5\us$ time steps. Both ratios are layer-by-layer MSE divided by terminal MSE for the same prescribed evolution. The ideal input has $\langle X\rangle=0.5$ and $\langle Y\rangle=0$ in the control frame. The exact-model comparison uses $10^5$ production shots. In the calibration comparison both methods use the same known nonwhite history, $40000$ shots for two-quadrature endpoint calibration and $60000$ for production. Controls and readout are ideal, and bath realizations or sampling blocks are independent.}
\label{tab:background-echo}
\begin{tabular}{cccccc}\toprule
$v\ (\mathrm{rad}/\mu\mathrm{s})$ & Control & $K_{\rm end}^2$ & $K_{\rm layer}^2$ & Exact-model MSE ratio & MSE ratio with calibration\\\midrule
$0.23224$ & Free & $57.08$ & $16969.76$ & $298.58$ & $170.46$\\
$0.23224$ & Echo & $9.202$ & $6282.74$ & $701.82$ & $415.98$\\
$0.047936$ & Free & $67.43$ & $84.77$ & $1.258$ & $1.146$\\
$0.047936$ & Echo & $9.202$ & $18.50$ & $2.039$ & $1.615$\\\bottomrule
\end{tabular}
\end{table*}

Table~\ref{tab:background-echo} shows how much PEC can still gain after refocusing. Appendix~\ref{app:endpoint-calibration} bounds the shots needed to calibrate the complex endpoint coherence. With calibration included, the terminal echo MSE at $100\us$ is $2.520\times10^{-4}$ for either assignment. The layer-by-layer MSE is $0.1048$ for the faster fluctuator and $4.070\times10^{-4}$ for the slower one. Their respective MSE ratios, $415.98$ and $1.615$, reflect the different histories of coherence loss and return before the same final echo. For the faster assignment, changing only the assumed imbalance to $p=0.9$ reduces the ratio to $1.52$. Merging adjacent time steps can also reduce the layer-by-layer cost, since the positive part in \Cref{eq:echo-local-cost} is then taken after summing increments. The terminal measurement uses only the contrast remaining at readout.

\section{Discussion}\label{sec:discussion}
Must the number of circuit executions required for quantum error
mitigation grow exponentially with circuit depth? At fixed accuracy,
worst-case lower bounds show that such growth can be unavoidable even
for local Markovian noise acting independently at each layer
\cite{Takagi2022,Takagi,Quek2024}. Experiments on superconducting
processors, however, have resolved noise correlations across gates
\cite{White,Agarwal}, and experiments with engineered dephasing have
directly observed coherence returning after an initial decay
\cite{Khurana2019}. Such information backflow raises a practical
question for PEC. How much should it cost to correct a loss that the
environment itself later reverses? This can be immediately seen from a harmonic example, where the
original measurement recovers the ideal expectation value at every
full period. Yet independent inversion on a grid resolving each
half-period incurs a shot overhead that grows exponentially with the
number of periods.

This paper uses this general observation, but turns it into a specific design principle for noise-aware QEM. The bath correlation function specifies how a fluctuation at one time is correlated with a later one. The ideal gates specify which register operators those fluctuations act on. Integrating these contributions together gives both the coherent shift to compensate and the net dissipative error in \Cref{eq:cumulative-dissipation}. A Pauli measurement compiler can use the resulting response to correct a single target \cite{Eddins}. The same integrated response could supply a microscopic noise model for a global postprocessing method such as Algorithmiq's tensor-network error mitigation \cite{FilippovTEM,FilippovScalability,FischerTEMHardware,AlgorithmiqTEMFunction}. For general baths the two-point function controls this construction through second order. The remainder in \Cref{eq:cumulative-bias} sets the accuracy available at the chosen duration.

The echo example in Table~\ref{tab:background-echo} illustrates why the timing of coherence loss matters for PEC. The two fluctuator frequencies compatible with the published Ramsey data give the same final echo contrast and hence the same terminal shot overhead. The faster fluctuator, however, produces more cycles of loss and recovery before readout, increasing the cost of layer-by-layer inversion. Measuring both Ramsey quadratures at successive times along the same echo sequence would test this prediction. 

An alternative starting point for PEC is the generalized quantum master equation, e.g., ~\cite{LiWangGQME2023}. It represents bath memory using auxiliary modes, reproduces the register dynamics through second order in the coupling, and gives a Kraus representation for short forward steps on the enlarged space. Propagating the auxiliary model to readout would give a register map from which to construct a signed terminal correction. 

Higher-order noise spectroscopy offers a second route. For centered stationary noise, the bispectrum supplies the three-point correlation entering the cubic Dyson term. When odd moments vanish, the fourth-order term involves the connected four-point spectrum (the trispectrum) and products of two-point correlations. For a quantum bath these spectra must retain the operator orderings sampled by the forward and backward evolutions. Spectroscopy protocols for non-Gaussian dephasing and experimental reconstruction of a bispectrum with a superconducting qubit provide a starting point \cite{NorrisPolyspectra2016,SungBispectrum2019}. 

The same distinction between temporary and persistent loss raises a question for quantum error correction. Liu et al.~\cite[Secs.~III and IV]{Liu} connect Pauli twirling of a process with memory to single-shot error correction under spatially local stochastic noise and construct PEC for the multitime process, and Endo et al.~\cite{Endo} relate syndrome-dependent recovery to the return of logical information. For a fixed code, decoder, and storage time, a natural question is whether the bath correlations can predict a spacing of syndrome measurements that lowers the final logical error. Answering this question requires following the encoded register and bath through syndrome extraction and feedback, since each measurement conditions their subsequent evolution. Comparing the decoded logical error at each spacing would test whether the predicted recovery persists through noisy syndrome measurements.

\section{Acknowledgements}

\subsection{AI Use Disclosure}
The author used OpenAI's ChatGPT-6.2  to assist in deriving
analytical expressions, writing code for the numerical experiments,
and cross-checking the analysis.  The author verified these outputs and takes
responsibility for the results reported in this paper.

\subsection{Funding support}
The work is supported by NSF Grant DMS-2552687, and a Superseed grant from Penn State Institute for Computational and Data Sciences (ICDS).

\subsection{Data availability}
The source package accompanying this draft contains the numerical scripts, model parameters, generated data, and figures. All new results are classical calculations. The published Ramsey data used as conditional input are from Oda et al.\ \cite{Oda,OdaData}, release v0.0.2 under CC BY 4.0. 
\appendix

\section{Exact costs for scalar dephasing}\label{sec:scalar}

Consider first a single qubit undergoing dephasing in the $Z$ basis. The channel leaves $I$ and $Z$ unchanged and attenuates $X$ and $Y$ by the same coherence factor $e^{-\chi(t)}$. The same description applies to a register when a Hermitian operator $Q$ with eigenvalues $+1$ and $-1$ defines the two dephasing sectors, as for $Q=Z_1Z_2$. We call an operator $A$ even if $QAQ=A$ and odd if $QAQ=-A$. Even operators act within the eigenspaces of $Q$, while odd operators connect the two eigenspaces. We assume that the channel fixes the even operators and multiplies the odd ones by $e^{-\chi(t)}$, with $\chi(0)=0$ and $0\leq\chi(t)<\infty$.

A correction $u\id+v\Ad_Q$ acts by $u+v$ on the even operators and by $u-v$ on the odd ones. For an exponent increment $\Delta_j=\chi(t_j)-\chi(t_{j-1})$, inversion therefore requires $u+v=1$ and $u-v=e^{\Delta_j}$. These conditions fix the weights and give the local normalization $|u|+|v|=e^{[\Delta_j]_+}$. An interval of coherence loss costs $e^{\Delta_j}$ to invert. During a recovery interval the local normalization is one, so independent sampling retains the cost incurred during the earlier loss.

If $Q\otimes I_B$ commutes with the joint evolution in this frame, interior $Q$ gates can be moved to the endpoint~\cite{Hakoshima}. The net correction acting on the noisy output is $\id$ for an even number of $Q$ gates and $\Ad_Q$ for an odd number. Summing the signed coefficients of schedules with the same parity before sampling gives the terminal inverse
\begin{equation}\label{eq:compressed-map}
 R_T=\frac{1+e^{\chi(T)}}{2}\id
     +\frac{1-e^{\chi(T)}}{2}\Ad_Q.
\end{equation}
The two procedures therefore have exact normalizations
\begin{equation}\label{eq:compression-costs}
 \begin{aligned}
 K_{\mathrm{layer}}&=e^{\sum_j[\Delta_j]_+},\qquad
 K_{\mathrm{end}}=e^{\chi(T)},\\
 \frac{K_{\mathrm{layer}}^2}{K_{\mathrm{end}}^2}
 &=\exp\!\left(2\sum_j[-\Delta_j]_+\right).
 \end{aligned}
\end{equation}
For a single qubit, $K_{\mathrm{end}}$ agrees with the exact inverse-dephasing cost of Ref.~\cite[Lemma~15]{Jiang}.

Here the recovery bound in \Cref{eq:price-endpoint} is attained by a pair of states of the register alone. Choose eigenstates $\lvert q_+\rangle$ and $\lvert q_-\rangle$ of $Q$ with opposite eigenvalues. The two superpositions $(\lvert q_+\rangle\pm\lvert q_-\rangle)/\sqrt{2}$ have trace distance $D_{\rm tr}(t)=e^{-\chi(t)}$ after dephasing. For absolutely continuous $\chi$ on $[0,T]$, the fine-grid limit of \Cref{eq:compression-costs} gives
\begin{equation}\label{eq:log-backflow}
 \log\frac{K_{\mathrm{layer}}}{K_{\mathrm{end}}}
 =\int_0^T\left[\frac{d}{dt}\log D_{\rm tr}(t)\right]_+dt.
\end{equation}
The trace-distance measure integrates increases in $D_{\rm tr}$~\cite{BLP2009}. For a single qubit with generator $\gamma(t)(Z\rho Z-\rho)$, $\dot\chi=2\gamma$ and \Cref{eq:log-backflow} equals $2\int_0^T[-\gamma(t)]_+dt$, the unnormalized Rivas--Huelga--Plenio measure with a normalized Choi state~\cite{RHP,Hall,Hakoshima}. Thus the negative-rate integral has an operational value here. It equals the excess log normalization paid by independent sampling on a grid resolving the recovery. For a Pauli observable that anticommutes with $Q$, the terminal correction is simply to multiply the noisy outcome by $e^{\chi(T)}$. At a full revival this factor is one, and the original measurement gives the ideal expectation value.

\section{Computing the layer-by-layer and terminal PEC benchmarks}\label{app:price-proof}
Section~\ref{sec:gross-net} defines the optimal normalizations
$K_{\mathrm{layer}}$ and $K_{\mathrm{end}}$ in \Cref{eq:optimal-costs}.
Here we give the semidefinite program used in
Sec.~\ref{sec:numerics-explicit} to compute both benchmarks from the
exact reduced channels of the retained-bath model.

For a Hermiticity-preserving, trace-preserving map $\mathcal F$, write $J(\mathcal F)=(\id\otimes\mathcal F)(\ketbra{\Omega})$ with $\ket{\Omega}=\sum_{a=1}^d\ket{aa}$. Grouping the positive and negative coefficients of a signed channel decomposition gives the program~\cite{Jiang,Regula2021}
\begin{align}
 \text{minimize }&\quad \mu_++\mu_-\nonumber\\
 \text{subject to }&\quad J_+-J_-=J(\mathcal F),\quad J_\pm\succeq0,\nonumber\\
 &\quad \Tr_{\rm out}J_\pm=\mu_\pm I,\quad \mu_\pm\ge0 .
 \label{eq:exact-pec-sdp}
\end{align}
Solving \Cref{eq:exact-pec-sdp} for each $\mathcal F_j=\Phi_{j-1}\Phi_j^{-1}$ and multiplying the optima gives $K_{\rm layer}$. One solve for $\Phi_L^{-1}$ gives $K_{\rm end}$. Each of these $L+1$ programs uses Choi matrices of size $d^2\times d^2$.

\section{Cumulative PEC and terminal error guarantees}\label{app:control-proofs}
The two estimators in Sec.~\ref{sec:control} use the same integrated bath response. Here we derive their bias bounds in Theorem~\ref{thm:cumulative-cost} and show how the bath spectrum controls the two-channel normalization.
\subsection{Remainder in the microscopic expansion}\label{app:tcl-derivation}\label{app:dyson}
The fixed-duration bias guarantee uses the remainder $\rho_3(T)$ in \Cref{eq:remainder-certificate}. For a centered, initially factorized bath, the first Dyson term vanishes after tracing out the bath. The second term is the ordered double integral in \Cref{eq:cumulative}~\cite{breuer2002theory,WangLi}.
For bounded $V(t)$, the $k$th iterated Duhamel remainder contains $k$ commutators and one exact joint unitary propagator. Each commutator has diamond norm at most $2|\lambda|\norm{V(t)}_\infty$. The propagator, bath preparation, and partial trace have diamond norm one. Integration over the ordered simplex bounds this remainder by $a_T^k/k!$, where
\[
 a_T=2|\lambda|\int_0^T\norm{V(t)}_\infty\,dt .
\]
Taking $k=3$ gives the certificate $\rho_3(T)=a_T^3/3!$. If the third reduced coefficient vanishes, taking $k=4$ gives $\rho_3(T)=a_T^4/4!$.

When a finite joint Hamiltonian is specified, the same estimate gives a tighter check on a particular terminal measurement. Let $\Phi_{<p}$ denote the laboratory-frame reduced map obtained by truncating the interaction-picture Dyson expansion at degree $p-1$ in $\lambda$ and restoring the ideal output rotation. For a laboratory observable $M$ and ideal input-frame target $B$, add and subtract $\Phi_{<p}^\dagger(M)$ in \Cref{eq:terminal} to obtain
\begin{equation}\label{eq:finite-bath-validation}
 \norm{\Phi_T^\dagger(M)-B}_\infty
 \le \norm{\Phi_{<p}^\dagger(M)-B}_\infty
     +\norm{M}_\infty\frac{a_T^p}{p!}.
\end{equation}
The first term retains cancellations through degree $p-1$. We compute it by extending the triangular block propagation in Appendix~\ref{app:retained-time} to $p$ blocks. This uses the specified joint model beyond its two-point correlations to validate the original estimator. The measured observable and its sampling weights are those of the second-order construction.

\subsection{Bounds for the cumulative dissipator}\label{app:cumulative-proof}
For the Hermitian integrated bath operators $Y_a$ in Sec.~\ref{sec:control}, \Cref{eq:cumulative-gram} gives $D_{ab}=\lambda^2\Tr(\tau_B Y_bY_a)$. For any vector $v$, $v^\dagger Dv=\lambda^2\Tr(\tau_B Y_v^\dagger Y_v)\ge0$, where $Y_v=\sum_a\overline{v_a}Y_a$. Thus the jump map $\mathcal J_D$ is completely positive. The half-anticommutator with $R_D$ has diamond norm at most $r_T$, giving
\[
 \norm{\mathcal J_D}_\diamond
 =\norm{\mathcal J_D^\dagger(I)}_\infty=r_T,\qquad
 \norm{\cD_D}_\diamond\le2r_T .
\]
Orthonormality of the $F_a$ gives $\Tr R_D=\Tr D$.

For $r_T>0$, the Kraus operators in \Cref{eq:completion-channel} obey $(\sum_\mu L_\mu^\dagger L_\mu+W_T^2)/r_T=I$. Expanding $r_T(\mathcal C_T-\id)$ gives $\cD_D+\cD_{W_T}$, proving \Cref{eq:inverse-completion}. At $r_T=0$ all jump operators vanish and the remaining correction is unitary. The same positivity argument applied to the instantaneous rate matrices gives $0\preceq R_D\preceq\int_0^T R_+(t)\,dt$. Taking operator norms yields the comparison in \Cref{eq:cumulative-versus-local}.

\subsection{Cost and bias of the terminal corrections}\label{app:pauli-sampling}
To prove the Pauli-readout part of Theorem~\ref{thm:cumulative-cost}, substitute the observable from \Cref{eq:linear-observable-compiler} into the adjoint of the physical channel. The second-order terms cancel, leaving
\[
 \widetilde\Phi_T^\dagger(O_T^{(2)})-B
 =-(\mathcal K_2^\dagger)^2 B
    +\mathcal R_3^\dagger(\id-\mathcal K_2^\dagger)B .
\]
Taking the operator norm proves \Cref{eq:linear-observable-bias}. For a Pauli target, expand the output correction as $\Delta O=\overline O_T^{(2)}-O=h_I I+\sum_{Q\ne I}h_Q Q$. Its identity coefficient is a known offset. Pauli orthogonality and $\|\Delta O\|_\infty\le\kappa_T$ give
\[
 \sum_{Q\ne I}|h_Q|^2
 =2^{-n}\Tr[(\Delta O)^2]-|h_I|^2\le\kappa_T^2 .
\]
If $s_T$ counts the nonzero coefficients with $Q\ne I$, Cauchy--Schwarz gives $K_{\rm meas}\le 1+\sum_{Q\ne I}|h_Q|\le 1+\sqrt{s_T}\,\kappa_T$. With $m$ Pauli couplings and $L$ piecewise-constant Clifford frames, each pair of coupling and interval indices produces a constant number of Pauli products. Consequently $s_T=O(m^2L^2)$.

For the two-channel correction let $b=(1+r_T)^{-1}$. The Taylor remainder of the CPTP semigroup is $\int_0^b(b-u)e^{u\mathcal L_{0,T}}\mathcal L_{0,T}^2\,du$. Substitution into \Cref{eq:two-branch-correction} gives
\begin{equation}\label{eq:two-branch-remainder}
 \mathcal R_T^{(2)}=\id-\mathcal K_2+\mathcal E_4,\qquad
 \norm{\mathcal E_4}_\diamond\le\frac{\ell_T^2}{2(1+r_T)} .
\end{equation}
Composition with the physical channel gives
\[
 \mathcal R_T^{(2)}\widetilde\Phi_T-\id
 =-\mathcal K_2^2+(\id-\mathcal K_2)\mathcal R_3
    +\mathcal E_4\widetilde\Phi_T .
\]
Since $\norm{\widetilde\Phi_T}_\diamond=1$, the triangle inequality proves \Cref{eq:cumulative-bias}. The two absolute coefficients sum to $1+2r_T$.

For either measurement protocol, let $\epsilon_{\rm fit}$ be the modeled bias and let $\epsilon_s$ bound the deviation of the actual observable response of branch $s$ from its model. The mean bias increases by at most $\sum_s|w_s|\epsilon_s$. Each random score lies in $[-K,K]$ after subtracting any known offset. Hoeffding's inequality gives, with probability at least $1-\delta$ for $N$ independent shots,
\begin{equation}\label{eq:terminal-bound}
 |\widehat\mu-\Tr(\rho_0B)|\le\epsilon_{\rm fit}
 +\sum_s|w_s|\epsilon_s
 +K\sqrt{\frac{2\log(2/\delta)}{N}} .
\end{equation}

For the optional random-unitary implementation of $\mathcal P_T$, diagonalize $W_T$. Averaging $e^{-i\theta W_T}Xe^{i\theta W_T}$ damps its $jk$ entry by $e^{-b(w_j-w_k)^2/2}$, which is precisely $e^{b\cD_{W_T}}$ in this basis. Duhamel's formula for the product of the two semigroups gives
\[
 \norm{e^{b\mathcal H}e^{b\cD_{W_T}}-e^{b(\mathcal H+\cD_{W_T})}}_\diamond
 \le \tfrac12 b^2\norm{[\mathcal H,\cD_{W_T}]}_\diamond .
\]
Indeed, expanding the difference as a double integral inserts one commutator between CPTP propagators of norm one. The positive branch has weight $1/b$, giving the stated added bias.

\subsection{Calibration error}
Algorithm~\ref{alg:observable-compiler} accepts a bound $\Delta_{\rm model}$ on the error in the integrated generator. To compute such a bound from fitted parameters $\widehat D\succeq0$ and $\widehat H_2=\widehat H_2^\dagger$, let
\begin{equation}\label{eq:cumulative-calibration}
 \Delta_{\rm cal}=2\norm{H_2-\widehat H_2}_\infty+2\norm{D-\widehat D}_1 .
\end{equation}
Splitting $D-\widehat D$ into its positive and negative parts and using $\norm{\cD_D}_\diamond\le2\Tr D$ for positive $D$ gives $\norm{\mathcal K_2-\widehat{\mathcal K}_2}_\diamond\le\Delta_{\rm cal}$. The physical channel therefore has a linear expansion about $\id+\widehat{\mathcal K}_2$ with residual bounded by $\rho_3(T)+\Delta_{\rm cal}$. Substitution into \Cref{thm:cumulative-cost} gives the calibrated bounds with the fitted $\kappa_T$, $\ell_T$, and $r_T$. Duhamel's formula between the two CPTP semigroups also gives $\norm{e^{\mathcal K_2}-e^{\widehat{\mathcal K}_2}}_\diamond\le\Delta_{\rm cal}$.

\subsection{Spectral bound for the cumulative cost}
To obtain the spectral cost bound in \Cref{eq:general-filter-budget}, we express $R_D$ in terms of the integrated interaction $V_T=\int_0^T V(t)\,dt$. Substituting $V_T=\sum_a F_a\otimes Y_a$ into the Gram representation gives
\begin{equation}\label{eq:integrated-interaction-radius}
 R_D=\lambda^2\Tr_B[(I\otimes\tau_B)V_T^2].
\end{equation}
For unbounded bath operators the expectation is understood as a quadratic form with finite second moments. Inserting the bath spectrum and the filters of \Cref{eq:operator-filter} yields
\begin{align}
 R_D
 &=\lambda^2\sum_{\alpha\beta}\int_0^Tdt\int_0^Tds\,
 C_{\alpha\beta}(t-s)\widetilde A_\alpha(t)\widetilde A_\beta(s)\nonumber\\
 &=\lambda^2\int\sum_{\alpha\beta}
 [J(\omega)]_{\alpha\beta}
 \widehat A_{\alpha,T}(\omega)^\dagger\widehat A_{\beta,T}(\omega)\,d\omega .
 \label{eq:operator-filter-component}
\end{align}
In the fixed operator basis put $[F_T^{\rm op}(\omega)]_{a\alpha}=\int_0^T e^{i\omega t}a_{a\alpha}(t)\,dt$. The reversed bath indices in \Cref{eq:cumulative-gram} give
\begin{equation}\label{eq:cumulative-filter}
 D(T)=\lambda^2\int F_T^{\rm op}(\omega)J(\omega)^{\mathsf T}
                       F_T^{\rm op}(\omega)^\dagger\,d\omega .
\end{equation}
The positive integrand $S_T(\omega)$ in \Cref{eq:operator-filter-radius} satisfies $\norm{S_T(\omega)}_\infty\le\Tr S_T(\omega)$. Integration and $\Tr R_D=\Tr D$ yield \Cref{eq:general-filter-budget}.

For \Cref{prop:shifted-moments}, positivity of the spectrum gives $\norm{S_T(\omega)}_\infty\le\norm{J(\omega)}_\infty\sum_\alpha\norm{\widehat A_{\alpha,T}(\omega)}_\infty^2$. Expand each filter into its Bohr-frequency components and use $|F_T(\omega+\nu)|\le2/|\omega+\nu|$. The triangle inequality in $L^2(\norm{J(\omega)}_\infty\,d\omega)$ gives
\[
 r_T\le4\lambda^2\sum_\alpha
 \left(\sum_{\nu\in V_\alpha}\norm{A_{\alpha\nu}}_\infty\sqrt{B_\nu}\right)^2 ,
\]
as required. Discrete spectral lines give the same bound with weighted sums.

\subsection{Exact inversion of the cumulative model}
The middle cost $K_{\rm cum}$ in \Cref{eq:cumulative-versus-local} isolates the effect of integrating the noise before decomposing its inverse. We obtain it by inverting the cumulative model exactly and then bounding the error from using that model for the physical evolution.
\begin{theorem}[Cost of inverting the cumulative model]\label{thm:poisson}
The signed expansion of $e^{-\mathcal K_2(T)}$ obtained from \Cref{eq:inverse-completion} has normalization $K_{\rm cum}=e^{2r_T}$. For $\eta_T=\norm{\widetilde\Phi_T-e^{\mathcal K_2(T)}}_\diamond$, applying this inverse to the physical channel gives bias at most $K_{\rm cum}\eta_T$, with
\begin{equation}\label{eq:cumulant-model-error}
 \eta_T\le\rho_3(T)+\tfrac12\kappa_T^2 .
\end{equation}
\end{theorem}
To see this, take $n$ Poisson with mean $r_T$. Conditional on $n$, draw $n$ sorted uniform points in $[0,1]$. Let $\mathcal G_{\boldsymbol u}$ be the CPTP composition of $n$ copies of $\mathcal C_T$ separated by drift channels $e^{v\mathcal L_{0,T}}$, with $v$ equal to each successive spacing including the two endpoints. The conditional density $n!$ on the ordered simplex turns its Dyson series into
\begin{equation}\label{eq:poisson-inverse}
 e^{2r_T}\mathbb E[(-1)^n\mathcal G_{\boldsymbol u}]
 =e^{r_T}e^{\mathcal L_{0,T}-r_T\mathcal C_T}=e^{-\mathcal K_2(T)} .
\end{equation}
For $r_T=0$ this expression reduces to the coherent inverse. Thus $\norm{e^{-\mathcal K_2}}_\diamond\le e^{2r_T}$. The remaining error follows from $\norm{e^{\mathcal K_2}-\id-\mathcal K_2}_\diamond\le\kappa_T^2/2$ by the integral Taylor formula for the forward CPTP semigroup.

For distinct nonidentity Pauli strings with $\cD_D=\sum_k b_k(\Ad_{P_k}-\id)$ and $b_k\ge0$, $R_D=(\sum_k b_k)I$ gives $r_T=\sum_k b_k$. The linear dissipative inverse $(1+r_T)\id-\sum_k b_k\Ad_{P_k}$ has optimal normalization $1+2r_T$ by orthogonality of the Pauli Choi vectors~\cite{Jiang}. Independent Pauli--Lindblad factors for the exponential inverse have normalization $e^{2r_T}$ before repeated Pauli strings are combined~\cite{VanDenBerg}.

\section{Analytical separations for noncommuting couplings}\label{app:many-proof}
The harmonic dephasing example in Sec.~\ref{sec:spectral} separates
the costs of independent inversion and a single terminal correction.
Here we examine the same comparison for noncommuting couplings. The
first model fixes the duration and varies the number of qubits. The
second fixes the coupling strength and varies the number of periods.

\subsection{A driven register at fixed duration}\label{sec:many}
We first apply the two-channel construction of Sec.~\ref{sec:control} to a driven register. It gives a deterministic correction whenever the integrated dissipator vanishes. To determine the residual bias and compare with the exact inverse costs in \Cref{eq:optimal-costs}, consider $M$ independent qubits driven for $T=2\pi$ by quasistatic transverse fields. The Hamiltonian of each qubit is
\begin{equation}\label{eq:manymodel}
 H_\xi=\tfrac12(Z+\lambda\xi X),\qquad H_{\rm id}=Z/2 .
\end{equation}
The fields $\xi\sim\mathcal N(0,1)$ are independent across qubits and shots. The interaction-frame
coupling is $(X\cos t-Y\sin t)/2$. In the normalized Pauli basis,
the cumulative coefficients in \Cref{eq:cumulative-gram} are
\[
 \begin{aligned}
 v(t)&=(\sin t,\cos t-1,0)^T/\sqrt2,\\
 D(t)&=\lambda^2v(t)v(t)^\dagger,\qquad
 H_2(t)=\tfrac{\lambda^2}{4}(t-\sin t)Z .
 \end{aligned}
\]
Thus $D(2\pi)=0$, and the two-channel correction in
\Cref{eq:two-branch-correction} reduces to the inverse coherent rotation.

\begin{theorem}[Fixed-time separation]\label{thm:many}
For \Cref{eq:manymodel} and $0<\lambda\le0.1$, coherent compensation
has normalization one and diamond-norm bias at most $18M\lambda^4$.
Given the exact endpoint channel, a tensor product of signed unitary
corrections removes the residual bias with normalization
$K_{\rm end}\le e^{15M\lambda^4}$. On the grid $0,\pi,2\pi$, every
independent exact full-channel inverse has
$K_{\rm layer}\ge e^{M\lambda^2}$, including decompositions over joint
$M$-qubit CPTP corrections.
\end{theorem}

\begin{proof}
For one qubit set $u=\lambda^2\xi^2$, $h=\sqrt{1+u}$ and
$\delta=2\pi(h-1)$. Averaging the two signs of $\xi$ at fixed $u$ gives
the Bloch matrix
\begin{equation}\label{eq:period-fixed-u}
 M_u=
 \begin{pmatrix}
 \frac{u+\cos\delta}{1+u}&-\frac{\sin\delta}{h}&0\\[2pt]
 \frac{\sin\delta}{h}&\cos\delta&0\\[2pt]
 0&0&\frac{1+u\cos\delta}{1+u}
 \end{pmatrix}.
\end{equation}
Compare this map with a $Z$ rotation of angle $\theta=\pi u$.
Their difference is the sum of a map on the transverse coherences,
of diamond norm $|z_u|$, and $e_u(\Ad_X-\Ad_Z)/2$, with
\[
 \begin{aligned}
 z_u&=\cos\delta-\cos\theta+i(\sin\delta/h-\sin\theta),\\
 e_u&=u(1-\cos\delta)/(1+u).
 \end{aligned}
\]
The inequalities $0\le\theta-\delta=\pi(h-1)^2\le\pi u^2/4$ and
$1-1/h\le u/2$ give $|z_u|\le3\pi u^2/4$ and
$e_u\le\pi^2u^3/2$. After the inverse rotation of angle $\pi\lambda^2$,
the comparison channel has coherence factor
$k=\mathbb E e^{i\pi\lambda^2(\xi^2-1)}$.
Centering the phase gives $|1-k|\le\pi^2\lambda^4$.
Since $\mathbb E u^2=3\lambda^4$ and $\mathbb E u^3=15\lambda^6$,
the one-qubit bias is at most
\[
 \left(\pi^2+\frac{9\pi}{4}+\frac{15\pi^2\lambda^2}{2}\right)
 \lambda^4\le18\lambda^4 .
\]
Telescoping the tensor product proves the first assertion.

Write $c=\mathbb E\cos\delta$, $b=\mathbb E(\sin\delta/h)$,
$e=\mathbb E e_u$ and $q=\sqrt{(c+e/2)^2+b^2}$. The singular values
of $\mathbb E M_u$ are $s_1=q+e/2$, $s_2=q-e/2$ and $s_3=1-e$.
Here $c\ge1-3\pi^2\lambda^4/2>0$ and $s_2\ge c$. The determinant
is positive, so input and output qubit rotations diagonalize the map.
In these rotated frames, the signed Pauli coefficients of the inverse
are given by
\[
 \begin{aligned}
 4q_0&=1+s_1^{-1}+s_2^{-1}+s_3^{-1},\\
 4q_j&=1+s_j^{-1}-s_k^{-1}-s_l^{-1},
 \end{aligned}
\]
where $\{j,k,l\}=\{1,2,3\}$.
The relation $s_2=s_1+s_3-1$ and $s_1,s_3\le1$ give
$q_1,q_3\le0$ and
$4q_2=(1-s_1)(1-s_3)(s_1+s_3)/(s_1s_2s_3)\ge0$.
The one-qubit normalization therefore satisfies
\[
 \sum_{j=0}^3|q_j|=s_2^{-1}
 \le(1-3\pi^2\lambda^4/2)^{-1}\le e^{15\lambda^4}.
\]
Taking the tensor product of these decompositions yields the terminal bound.

At $t=\pi$ the exact channel maps $Z$ to $a_\lambda Z$, where
$a_\lambda=\mathbb E[(1+u\cos(\pi h))/(1+u)]$.
For all $u\ge0$, $u-u^2\le u/(1+u)\le u$ and
$1-\cos(\pi(h-1))\le\pi^2u^2/8$. In the integrand for $1-a_\lambda$, write
\[
 \frac{u}{1+u}\bigl[2-(1-\cos(\pi(h-1)))\bigr]
 \ge 2(u-u^2)-\frac{\pi^2u^3}{8}.
\]
Averaging gives
\[
 1-a_\lambda\ge2\lambda^2-6\lambda^4
                 -15\pi^2\lambda^6/8\ge\lambda^2 .
\]
The single-qubit centered Duhamel bound
$\norm{\widetilde\Phi_t-\id}_\diamond\le\lambda^2t^2/2<1$
ensures invertibility up to $2\pi$ and $a_\lambda>0$.
For the input pair $\rho_\pm=(I\pm Z^{\otimes M})/2^M$, the first
inverse factor restores distinguishability by $a_\lambda^{-M}$.
Trace-norm contraction of each CPTP branch gives
$K_{\rm layer}\ge a_\lambda^{-M}\ge e^{M\lambda^2}$.
\end{proof}

\subsection{An exact recurrence in a finite bath}\label{sec:net-rate}\label{sec:finite-revival}
We now apply the distinguishability bound of Sec.~\ref{sec:gross-net}
to a single qubit with a finite, non-Gaussian bath. A transverse
component contracts by the same factor in each period and recovers
fully at readout. Following this component gives a lower bound on the
independent-inversion cost for arbitrarily many periods.
Take $H_{\rm id}=0$ and $H_\xi=\lambda\xi\cdot\sigma/2$.
On each shot choose a static field from $0,\pm gn,\pm2gm$, with
probability $q$ at zero and $(1-q)/4$ at each nonzero value.
For $\omega=\lambda g>0$ write $\tau=2\pi/\omega$ and
$D_v(c)=cI+(1-c)vv^T$. Averaging opposite rotations gives the exact
Bloch matrix
\begin{equation}\label{eq:finite-bloch-map}
 \mathsf T_t=qI+\tfrac{1-q}{2}
 [D_n(\cos\omega t)+D_m(\cos2\omega t)] .
\end{equation}
Fix $q=3/4$, $n=e_z$ and $m=(e_x+e_z)/\sqrt2$.
Every map is invertible since $\mathsf T_t\succeq(2q-1)I$.
At one quarter and one half of a period the matrices are
\[
 \begin{aligned}
 \mathsf T_{\tau/4}
 &=\tfrac18\begin{pmatrix}6&0&1\\0&5&0\\1&0&7\end{pmatrix},\\
 \mathsf T_{\tau/2}&=\operatorname{diag}(3/4,3/4,1).
 \end{aligned}
\]
They fail to commute since $[\mathsf T_{\tau/4},\mathsf T_{\tau/2}]_{xz}=1/32$. At each full period
$\mathsf T_{N\tau}=I$, and the register returns to its input state.

On the grid $0,\tau/4,\tau/2,\tau$ repeated for $N$ periods, the
quarter-period step contracts the $y$ component by $5/8$ in each
period. Its inverse therefore costs at least $8/5$ by trace-norm
contraction. All other inverse factors cost at least one. Hence
\begin{equation}\label{eq:noncommuting-separation}
 K_{\rm layer}(N\tau)\ge(8/5)^N,\qquad K_{\rm end}(N\tau)=1 .
\end{equation}
Both procedures are feasible in the same single-qubit Clifford library.
Its 24 Bloch rotations are the signed permutation matrices of determinant
one, and their average is zero. For each permutation, summing its four
allowed row-sign patterns with weights equal to the sign of row $i$
isolates the matrix unit in that row. These weighted sums span all real
$3\times3$ matrices, and the zero average allows the coefficients to be
chosen with sum one.

\section{Conditional terminal measurement}\label{app:readout-proof}
The spectral criterion in Sec.~\ref{sec:spectral} uses the terminal
score radius $K_P(T)$ in \Cref{eq:readout-radius-main}. We show how
to attain this radius with a measurement conditioned on the observed
pair and why it is optimal among single-copy readouts that are unbiased for
every input state.

For a Pauli target $P$ with nonempty bit-flip set $J=\mathcal J_P$,
let $\Pi_u$ project onto the span of the two basis states in pair
$u=\{z,z'\}$. Choose an ordering and put $k_u=k_T(z,z')$. By
\Cref{eq:schur}, the adjoint equation fixes the $(z,z')$ entry of
the corrected observable at $P_{zz'}/\overline{k_u}$. When every contributing
coherence factor $k_u$ is nonzero, these two-dimensional blocks give
\begin{equation}\label{eq:readout-moments}
 \widetilde\Phi_T^\dagger(O_T)=P,\qquad
 O_T^2=\sum_u |k_u|^{-2}\Pi_u .
\end{equation}
If $Y$ is the measured eigenvalue of $O_T$ and
$p_u=\Tr(\rho_0\Pi_u)$ is the conserved pair population, then
\[
 \mathbb E Y=\mu=\Tr(\rho_0P),\qquad
 \Var(Y)=\sum_u p_u|k_u|^{-2}-\mu^2.
\]
Since $|Y|\le K_P(T)$, Hoeffding's inequality gives statistical error
at most $\epsilon$ with confidence $1-\delta$ for
\begin{equation}\label{eq:readout-shots}
 N\geq 2K_P(T)^2\epsilon^{-2}\log(2/\delta)
\end{equation}
independent shots.

To prove optimality, take the two states
$\rho_\pm=(\Pi_u\pm\Pi_uP\Pi_u)/2$ in a single pair.
Their ideal expectations of $P$ differ by two, and their noisy outputs
have trace distance $|k_u|$. For any single-copy readout with scores bounded by $R$,
the two mean scores therefore differ by at most $2R|k_u|$.
Unbiasedness therefore requires $R\ge|k_u|^{-1}$ on every contributing
pair, which proves \Cref{eq:readout-radius-main}. If $k_u=0$, the
two outputs coincide and the target is unrecoverable. For an invertible channel, unbiasedness fixes the first-moment operator of any readout $\{E_y\}$ to be $\sum_y y E_y=O_T$. Its second-moment operator satisfies
\[
 \sum_y y^2 E_y-O_T^2
 =\sum_y (yI-O_T)E_y(yI-O_T)\succeq0 .
\]
The spectral measurement of $O_T$ attains equality for every input state~\cite{Watanabe}.

\begin{algorithm}[t]
\caption{Conditional measurement of the corrected Pauli observable}\label{alg:readout}
Given the nonzero terminal factors $k_u$ and the interaction-frame target $P$, compile the following measurement in the physical output frame by the transformation in \Cref{eq:branch}.
\begin{enumerate}
\item Choose a pivot $j_0\in J$. Apply a CNOT from $j_0$ to each other qubit in $J$. The two strings in a pair now differ only at the pivot. Measure the other qubits in $Z$ to identify $u$.
\item Measure the pivot in the equatorial basis of its $O_T$ block and record the eigenvalue $\pm|k_u|^{-1}$.
\item Average the scores over independent shots, with $N$ chosen from \Cref{eq:readout-shots}.
\end{enumerate}
The circuit uses $|J|-1$ CNOTs before routing and output-frame compilation. A real positive kernel permits a fixed equatorial axis with a classical sign determined by $u$.
\end{algorithm}

For a complex kernel the pivot block takes the form
\[
 O_{T,u}=|k_u|^{-1}(\cos\theta_u\,X+\sin\theta_u\,Y).
\]
A readout without feedforward chooses $X$ or $Y$ with equal
probability. For an outcome $y=\pm1$, the scores are
$2y|k_u|^{-1}\cos\theta_u$ and $2y|k_u|^{-1}\sin\theta_u$,
respectively. This doubles the second moment in \Cref{eq:readout-moments}.
If all target pairs have the same positive attenuation, measuring the
original Pauli and rescaling attains $K_P(T)$ with local readout alone.

For the real Gaussian kernel of \Cref{eq:schur} and couplings $Z_j/2$,
the nonzero entries of each $d_u$ form a sign vector on $J$. Denoting
the corresponding principal submatrix of $G(T)$ by $G_J$, we obtain
\begin{equation}\label{eq:radius-quadratic}
 \log K_P^2=\max_{s\in\{\pm1\}^{|J|}}s^\dag G_Js
 \leq |J|\lambda_{\max}(G_J).
\end{equation}
The upper bound gives a shot budget without enumerating the pairs.
For diagonal $G_J$, every contributing pair has the same attenuation
and $O_T=e^{\Tr G_J/2}P$.

\section{Spectral bounds and calibration}\label{app:spectral-proof}
Section~\ref{sec:spectral} expresses the shot overhead for a Pauli target in terms of the coherences seen by that target. Here we use this expression to prove the spectral criterion, derive the growth rates in Table~\ref{tab:spectral-scaling}, and quantify the error when the coherence must be estimated.

\subsection{Proof of Theorem~\ref{thm:spectral}}
To establish the boundedness criterion for both continuous spectra and spectral lines, let $M$ be the finite positive matrix-valued spectral measure of the bath. Its projection onto a coherence pair $u$ is
\[
 \nu_u(d\omega)=\lambda^2 d_u^\dag M(d\omega)d_u .
\]
For a density, $\nu_u(d\omega)=j_u(\omega)d\omega$ as in \Cref{eq:spectral-moment}. Substituting the spectral representation into \Cref{eq:covariance} and applying Fubini's theorem gives the variance of the Gaussian phase difference,
\begin{equation}\label{eq:proof-spectral-variance}
 v_u(T)=d_u^\dag G(T)d_u=\int |F_T(\omega)|^2\nu_u(d\omega).
\end{equation}
By \Cref{eq:readout-radius-main}, $\log K_P(T)^2=\max_u v_u(T)$. A zero-frequency atom contributes $T^2\nu_u(\{0\})$, which makes the overhead unbounded. In the absence of such an atom, set $\mathfrak m_u=\int_{\omega\ne0}\omega^{-2}\nu_u(d\omega)$. The filter bound $|F_T(\omega)|^2\leq4/\omega^2$ gives $\sup_T v_u(T)\leq4\mathfrak m_u$. For the reverse estimate, Tonelli's theorem gives
\begin{equation}\label{eq:proof-spectral-average}
 \frac{1}{L}\int_0^L v_u(T)dT
 =2\int_{\omega\ne0}
 \frac{1-\sin(\omega L)/(\omega L)}{\omega^2}\nu_u(d\omega).
\end{equation}
Fatou's lemma gives $2\mathfrak m_u\leq\sup_Tv_u(T)$, including when the integral diverges. Taking the maximum over the finite set of pairs proves \Cref{eq:spectral-iff}. For a density with finite inverse-square moment,
\[
 v_u(T)=2\mathfrak m_u-2\int_{\mathbb R}\cos(\omega T)
                   j_u(\omega)\omega^{-2}\,d\omega.
\]
The cosine transform vanishes as $T\to\infty$ by the Riemann--Lebesgue lemma, proving \Cref{eq:spectral-density-limit}. A monochromatic pair of lines attains $4\mathfrak m_u$ as in \Cref{eq:harmonic-rate}, and smooth densities concentrated near those lines approach the same maximum. The lower constant is also sharp. For example,
\[
 \begin{aligned}
 j_u(\omega)&=\mathfrak m_u\omega^2e^{-\omega^2/2}/\sqrt{2\pi},\\
 v_u(T)&=2\mathfrak m_u(1-e^{-T^2/2}).
 \end{aligned}
\]
For this density $\sup_T v_u(T)=2\mathfrak m_u$.

Gaussianity enters when converting the variance to a coherence. The field $\xi(t)=\sqrt{2}\sigma\cos(\omega t+\phi)$ with uniform $\phi\in[0,2\pi)$ has the same BCF as \Cref{eq:harmonic-bath}. Averaging the phase factor over $\phi$ gives $k(T)=J_0(\sqrt{8a}\sin(\omega T/2))$. For $a\ge j_{0,1}^2/8$, where $j_{0,1}$ is the first positive zero of $J_0$, a transverse coherence vanishes at some readout time despite the finite inverse-square moment.

\subsection{Low-frequency growth and control filters}\label{app:filter-scope}
The power-law rows of Table~\ref{tab:spectral-scaling} follow by separating frequencies below and above $1/T$ in \Cref{eq:proof-spectral-variance}. Suppose $j_u(\omega)\le c_u|\omega|^p$ for $|\omega|\le\omega_0$ with $p>-1$. The contribution above $\omega_0$ is bounded by $4B_{u,\rm hi}$, where
\[
 B_{u,\rm hi}=\int_{|\omega|>\omega_0}\omega^{-2}\nu_u(d\omega).
\]
For $T\ge1/\omega_0$, integrating the bound $|F_T(\omega)|^2\le\min(T^2,4/\omega^2)$ on the two low-frequency intervals gives
\begin{align}
 v_u(T)&\le\left(\frac{2c_u}{p+1}+\frac{8c_u}{1-p}\right)T^{1-p}
              +4B_{u,\rm hi}, & -1<p<1,\label{eq:proof-lowfrequency-power}\\
 v_u(T)&\le c_u+8c_u\log(\omega_0T)+4B_{u,\rm hi},
              & p=1.\label{eq:proof-lowfrequency-log}
\end{align}
For $p>1$, \Cref{eq:spectral-iff} gives bounded cost. A matching lower bound near zero gives the power $T^{1-p}$ for $-1<p<1$ by restricting $\omega T$ to a fixed interval where $\sin^2(\omega T/2)$ stays positive. For $p=1$, integrating over a fixed fraction of each oscillation gives a harmonic sum. The same orders apply to the envelope $j_*=\max_u j_u$ in the table, since the number of pairs is finite and, for $w=|F_T|^2$,
\[
 \max_u\int w j_u\le\int w j_*\le\sum_u\int w j_u .
\]
These estimates give the corresponding orders of $\log K_P(T)^2$~\cite{Trushechkin}.

The refocusing sequence in Sec.~\ref{sec:background-echo} changes the filter while preserving the coupling eigenbasis. More generally, let $f_u(t)$ be the vector of coupling differences for pair $u$ under controls that preserve this basis. Its variance is
\begin{equation}\label{eq:proof-control-filter}
 \begin{aligned}
 v_u(T)&=\lambda^2\int F_u(\omega)^\dag M(d\omega)F_u(\omega),\\
 F_u(\omega)&=\int_0^T e^{i\omega t}f_u(t)dt .
 \end{aligned}
\end{equation}
For $f_u$ of bounded variation, integration by parts gives
\[
 |\omega|\,\|F_u(\omega)\|_2
 \le \|f_u(0)\|_2+\|f_u(T)\|_2+\operatorname{TV}(f_u).
\]
A uniform bound on the right-hand side gives a uniform variance bound whenever $\int\omega^{-2}\Tr M(d\omega)$ is finite.

For the harmonic line in \Cref{eq:harmonic-bath}, two sign reversals in the last half-period give an explicit refocusing sequence. At readout $T_n=(2n+1)\pi/\omega$, choose $\omega t_1=2n\pi+\pi/3$ and $\omega t_2=2n\pi+2\pi/3$. For the switching function $y=+1,-1,+1$ on the three successive intervals, the filter obeys
\[
 i\omega\int_0^{T_n} y(t)e^{i\omega t}dt
 =e^{i\omega T_n}-1+2(e^{i\omega t_1}-e^{i\omega t_2})=0 .
\]
With both spectral lines refocused, the original terminal measurement has $K_{\rm end}=1$.

\subsection{Direct endpoint calibration}\label{app:endpoint-calibration}
The two-quadrature protocol in Sec.~\ref{sec:background-echo} estimates the complex terminal coherence directly. The following worst-case bound gives the dependence of its calibration and production costs on the signal remaining at readout. A trusted single-qubit $+X$ preparation followed by $X,Y$ measurements determines $h=\langle X\rangle+i\langle Y\rangle=\overline{k_T(0,1)}$. Suppose $|h|\ge c>0$, $0<\epsilon\le1$, with independent shots and the same terminal channel in calibration and production. Given $m$ calibration shots per quadrature, the complex sample mean satisfies
\[
 |\widehat h-h|\le e_h,\qquad
 e_h=2\sqrt{\log(4/\delta_{\rm cal})/m}
\]
with probability $1-\delta_{\rm cal}$. Accept $|\widehat h|\ge c/2$ and measure the fitted $X$ observable
\[
 \frac{\operatorname{Re}(\widehat h)X+\operatorname{Im}(\widehat h)Y}
      {|\widehat h|^2},
\]
or its orthogonal equatorial axis for $Y$ with the same normalization. For $e_h\le c\epsilon/4$, acceptance follows and the observable bias is at most $e_h/|\widehat h|\le\epsilon/2$. The production scores have radius at most $2/c$. Applying Hoeffding's inequality with statistical tolerance $\epsilon/2$ and assigning equal failure probabilities to calibration and production gives
\begin{equation}\label{eq:endpoint-calibration-cost}
 \begin{aligned}
 N_{\rm cal}+N_{\rm prod}&=O\!\left(c^{-2}\epsilon^{-2}\log(1/\delta)\right),\\
 \delta&=\delta_{\rm cal}+\delta_{\rm prod}.
 \end{aligned}
\end{equation}
For the harmonic bath, $c=e^{-2a}$ suffices for all durations.

\subsection{Harmonic example in the Choi-channel representation}
\label{app:liu-harmonic}
To obtain the three normalizations compared in Sec.~\ref{sec:intro},
consider two successive half-periods of \Cref{eq:harmonic-bath}, with
one gate between them. The accumulated phases are
$\vartheta=2\lambda y/\omega$ and $-\vartheta$, with
$\mathbb E\vartheta^2=4a$. In the Choi-channel representation of
Liu et al.~\cite[Sec.~IV, Eqs.~(8) and (9)]{Liu}, each noise interval
acts on its own register and both registers see the same realization
of $\vartheta$. The channel is therefore
\[
 \mathcal C=\mathbb E_\vartheta\,\Ad_{e^{-i\vartheta A}},
 \qquad A=(Z\otimes I-I\otimes Z)/2 .
\]
For eigenstates of $A$ with eigenvalues $h,h'$, the Gaussian average
multiplies their coherence by $e^{-2a(h-h')^2}$. In particular, it
reduces the trace distance between
$(\ket{01}+\ket{10})/\sqrt{2}$ and $(\ket{01}-\ket{10})/\sqrt{2}$
from one to $e^{-8a}$. Trace-norm contraction of each physical channel
in a signed decomposition of $\mathcal C^{-1}$ thus requires
$K_{\mathrm{process}}\ge e^{8a}$.

Four product unitaries attain this bound. With $S=\operatorname{diag}(1,i)$
and $U=S\otimes S^\dagger=e^{-i\pi A/2}$, write
$\mathcal C^{-1}=\sum_{k=0}^3 q_k\Ad_{U^k}$, where
\[
 \begin{aligned}
 q_0&=(1+2e^{2a}+e^{8a})/4,\qquad q_1=q_3=(1-e^{8a})/4,\\
 q_2&=(1-2e^{2a}+e^{8a})/4 .
 \end{aligned}
\]
On each eigenvalue difference $g=0,\pm1,\pm2$, these coefficients give
$\sum_k q_k e^{-i\pi k g/2}=e^{2ag^2}$, the required inverse
multiplier. Their absolute values sum to $e^{8a}$.

In the sequential circuit, branch $k$ applies $S^k$ after the first
noise interval and $S^{-k}$ after the second. With an intervening identity,
the phase rotations commute and cancel for every $\vartheta$ and every
$k$. All four branches therefore implement the identity, and their
combined coefficient is $\sum_k q_k=1$. The dephasing exponent at the
three interval boundaries is $0,2a,0$, so \Cref{eq:compression-costs}
gives $K_{\mathrm{layer}}=e^{2a}$ on this time grid. An intervening
$X$ gate instead makes the phases add in the interaction frame and
attenuates a transverse coherence by $e^{-8a}$. The full Choi channel
retains the response to both gate choices.

\section{Analysis of the numerical examples}\label{app:numerical-methods}
We derive the uniform cost bound in \Cref{eq:retained-uniform-cost} and
specify the circuits and calibration protocols used in
Tables~\ref{tab:pauli-main} and~\ref{tab:background-echo}.

\subsection{Exact propagation and cost bound for the retained bath}
\label{app:explicit-compiler-method}\label{app:retained-time}
In Sec.~\ref{sec:numerics-explicit}, the exact reference channel follows
by evolving the joint Hamiltonian in \Cref{eq:explicit-bath} and tracing
out the bath at readout. To extract the second Dyson coefficient, set
$\mathbb L_0=-i[H_0,\cdot]$, $\mathbb V=-i[V,\cdot]$, and form
\begin{equation}\label{eq:numerical-dyson-block}
 \mathbb A=\begin{pmatrix}
 \mathbb L_0&0&0\\ \mathbb V&\mathbb L_0&0\\
 0&\mathbb V&\mathbb L_0
 \end{pmatrix}.
\end{equation}
The lower-left block of $e^{T\mathbb A}$ gives the coefficient of
$\lambda^2$ in the joint propagator. Bath preparation, partial trace,
and removal of the ideal system rotation give $\mathcal K_2(T)$ after
multiplication by $\lambda^2$. Extending the same block construction to
higher orders gives the independent bias check in
\Cref{eq:finite-bath-validation}. Conjugation by $Z_B$ reverses $V$ and
preserves $H_0$ and $\tau_B$, so all odd reduced coefficients vanish.

\paragraph{Sharp bound on the sampling normalization.}
Put $f_j(T)=\int_0^T e^{ijt}\,dt$. In the free joint eigenbasis, the
only frequencies in the interaction are $\pm1,\pm2,\pm3$. The block of
the integrated interaction that flips the bath spin and the resulting
rate operator are
\[
 \begin{aligned}
 A&=\begin{pmatrix}-0.35i f_2&f_3/2\\ f_1/2&0.35i f_2\end{pmatrix},\\
 R_D/\lambda^2&=0.8AA^\dagger+0.2A^\dagger A .
 \end{aligned}
\]
Both are $2\pi$-periodic. Write $x=\cos T$ and set
\[
 \begin{aligned}
 a(x)&=\tfrac{49}{400}(1-x^2)+\tfrac{1-x}{4}
          +\tfrac{1-4x^3+3x}{36},\\
 q(x)&=-\tfrac{49}{400}(1-x^2)+\tfrac{x-2x^2+1}{6}.
 \end{aligned}
\]
Here $\Tr(A^\dagger A)=2a$ and $|\det A|=|q|$. Since $A$ is traceless
and $2\times2$, $AA^\dagger+A^\dagger A=2aI$. The largest eigenvalue
of $R_D$ is therefore
\begin{equation}\label{eq:retained-radius-exact}
 r_T=\lambda^2\left(a+\tfrac35\sqrt{a^2-q^2}\right).
\end{equation}
As $a\ge |q|$, the arithmetic--geometric mean inequality applied to
$2(a+q)$ and $(a-q)/2$ gives $\sqrt{a^2-q^2}\le(5a+3q)/4$. Hence
\[
 \frac{r_T}{\lambda^2}\le \frac74 a+\frac9{20}q
 =\frac{37}{45}
  -\frac{(x+1)(7000x^2+4133x+3667)}{36000}.
\]
The quadratic has negative discriminant and positive leading coefficient. At $x=-1$ we have $a=5/9$
and $q=-1/3$, giving equality. Thus the maximum $37/45$ is attained
at $T=\pi\bmod 2\pi$, as used in \Cref{eq:retained-uniform-cost}.
For the layer-by-layer curve, the integral of $2r_+(t)/\lambda^2$ over
one period is numerically $4.17243629$, giving a mean rate of $0.664064$.

For the independent-inverse benchmark we evaluate
\Cref{eq:exact-pec-sdp} on the exact reduced channels at $t_j=j/2$
through $T=400$. For the time-homogeneous baseline in
Fig.~\ref{fig:explicit-compiler}(b), we fit a GKSL generator to the
exact channels at $t=j/4$, $j=1,\ldots,24$, by least squares in the
Frobenius norm. Applying its inverse at later times tests extrapolation
from the calibration window.

\subsection{Four-qubit circuit and measurement protocols}
\label{app:layered-test}
The 15- and 16-layer tests in Sec.~\ref{sec:layered-test} use consecutive
depths of the same circuit. During each interval of duration
$\Delta=1/2$, the four data qubits interact with the same bath qubit
through $E_\lambda=\exp[-i\Delta(I_S\otimes H_B+\lambda V)]$, with
$\lambda=0.12$ and
\begin{align}
 H_B&=Z_B,\qquad \tau_B=(I_B+0.6Z_B)/2,\nonumber\\
 V&=A_x\otimes X_B+A_y\otimes Y_B,\label{eq:layered-interaction}\\
 A_x&=0.5Z_0+0.45Z_1+0.35Z_2+0.3Z_3,\nonumber\\
 A_y&=0.25X_1+0.2X_3.\nonumber
\end{align}
Each interval is followed by one of the alternating ideal layers
\begin{align}
 G_0&=\mathrm{CNOT}_{0\to1}\mathrm{CNOT}_{2\to3}H_0H_2,\nonumber\\
 G_1&=\mathrm{CNOT}_{1\to2}\mathrm{CNOT}_{3\to0}S_1S_3H_1H_3 .
 \label{eq:layered-gates}
\end{align}
Products act from right to left, starting with $E_\lambda$ followed by
$G_0$. For the ideal circuit $U_L$ the output target is $U_LX_0U_L^\dagger$.
Algorithm~\ref{alg:observable-compiler} expands its corrected observable as
\[
 \begin{aligned}
 O_L&=U_L(\mathcal I-\mathcal K_2^\dagger)(X_0)U_L^\dagger
     =cI+\sum_{P\ne I}a_PP,\\
 a_P&=\Tr(PO_L)/16,\qquad c=\Tr(O_L)/16 .
 \end{aligned}
\]
For the two-channel sampler, set
$S(t)=\int_0^t e^{2is}U_{\rm id}(s)^\dagger(A_x-iA_y)U_{\rm id}(s)\,ds$.
The two bath populations give
$R_D(t)/\lambda^2=0.2S(t)^\dagger S(t)+0.8S(t)S(t)^\dagger$.

\paragraph{Readout statistics and calibration.}\label{app:observable-practice}
For the product input in Table~\ref{tab:pauli-main}, let $\mu_A$ be the
mean of $O_L$ on the noisy output and $\mu_{\rm id}$ the ideal target
mean. Sampling one Pauli per shot with probability $|a_P|/K_{\rm meas}$
and recording $c+K_{\rm meas}\operatorname{sgn}(a_P)y$ for outcome $y=\pm1$ gives
\[
 \begin{aligned}
 K_{\rm meas}&=\sum_{P\ne I}|a_P|,\\
 \operatorname{MSE}&=(\mu_A-\mu_{\rm id})^2+
          \frac{K_{\rm meas}^2-(\mu_A-c)^2}{N_{\rm prod}} .
 \end{aligned}
\]
The table also reports the largest absolute bias over all input states.

For the untwirled calibration comparison, both estimators use the same
$10^4$ calibration outcomes and $9\times10^4$ independent production
shots. Drawing $s=\pm1$ uniformly, preparing $(I+sX_0)/16$, and
multiplying the output Pauli outcome by $s$ gives mean transfer
\[
 d(g)=\Tr[X_0\widetilde\Phi_g(X_0)]/16,\qquad g=\lambda^2 .
\]
We clip the measured transfer to the range of $d(g)$ on $0\le g\le0.12$.
Scalar rescaling divides the production mean by this estimate. The
compiler inverts $d(g)$ to estimate the strength and constructs $O_L$
from it. The reported MSE averages the conditional squared bias and
production variance over the calibration outcomes.

Under independent full Pauli frames, an interval transports the bath
operator $B$ accompanying a data Pauli $P$ according to
\begin{equation}\label{eq:twirled-bath-sector}
 \mathcal E_P(B)=\tfrac1{16}\Tr_S[(P\otimes I_B)
             E_\lambda(P\otimes B)E_\lambda^\dagger].
\end{equation}
Iterating from $P=X_0$, $B=\tau_B$ through the Clifford layers and taking
the final bath trace gives the endpoint attenuation. Both twirled
estimators use 20000 calibration shots, an attenuation floor of $0.8$,
and 80000 production shots. Endpoint calibration uses the full circuit.
The sparse Pauli--Lindblad fit~\cite{VanDenBerg} divides the calibration
shots among six single-interval probes, one for each generator
$(Z_0,Z_1,Z_2,Z_3,X_1,X_3)$. Their fitted inverse attenuations enter
the 16-layer target with multiplicities $(6,8,6,6,8,8)$.

\subsection{Noise model and matched echo comparison}\label{app:realistic-background}
The two frequencies in Fig.~\ref{fig:background-echo} come from the
published $X$-quadrature Ramsey data for qubit 8 of Algiers, using the
50 nominal delays in $[0,200]\us$ from release v0.0.2~\cite{Oda,OdaData}.
The least-squares fit is
\begin{equation}\label{eq:published-ramsey}
 p_0(t)=\tfrac12[1+e^{-at}\cos(\beta t)\cos(\xi t)],
\end{equation}
with $a=0.01109698\,\mu\mathrm{s}^{-1}$ and
$(\beta,\xi)=(0.04793604,0.23224239)\,\mathrm{rad}/\mu\mathrm{s}$.
Exchanging $\beta$ and $\xi$ preserves the fit. We therefore use each
as a candidate fluctuator frequency $v$ in \Cref{eq:background-coherence}.
The synthetic comparison sets the fluctuator polarization to $p=0.6$
and uses $r=a$ for the white dephasing floor.

The independent Gaussian background uses the one-sided spectrum
$S_f(f)=A/f$ with $A=3.6\times10^5\,\mathrm{Hz}^2$ from
Ref.~\cite{Burnett2019} and the chosen band
$[f_l,f_h]=[10^{-5},10^{-2}]\,\mathrm{Hz}$. With $t$ in microseconds,
put $F_y(f)=\int_0^T y(t)e^{i2\pi f10^{-6}t}dt$. Then
\begin{equation}\label{eq:background-filter}
 \chi_{1/f,y}(T)=\tfrac12(2\pi10^{-6})^2
       \int_{f_l}^{f_h}(A/f)|F_y(f)|^2df .
\end{equation}
The quasistatic variance is $\sigma_\omega^2=(2\pi10^{-6})^2A
\log(f_h/f_l)=9.81746\times10^{-5}\,\mu\mathrm{s}^{-2}$.
For $|y|\le1$, applying $1-\cos x\le x^2/2$ inside the covariance
integral gives
\begin{equation}\label{eq:background-static-error}
 \left|\chi_{1/f,y}(T)-\tfrac12\sigma_\omega^2
       \left(\int_0^T y(t)dt\right)^2\right|
 \le \frac{\sigma_\omega^2(2\pi10^{-6}f_h)^2T^4}{24}.
\end{equation}
The exponent error is below $2.59\times10^{-11}$ through $T=200\us$.

\paragraph{Matched echo protocol.}\label{app:background-echo}
Both estimators in Table~\ref{tab:background-echo} use the same
$0.5\us$ time steps, the same nonwhite dephasing exponents $f_j$, and a
midpoint pulse for the echo runs. The table evaluates the finite-step
cost in \Cref{eq:echo-local-cost}. Its fine-step limit follows by pairing
times $u$ and $T-u$, where the dephasing rates are $r+f'(u)$ and
$r-f'(u)$. The sum of their positive parts is $2r+[|f'(u)|-r]_+$, so
\begin{equation}\label{eq:echo-variation}
 \begin{aligned}
 \log K_{\mathrm e,layer}&=rT+\int_0^{T/2}[|f'(u)|-r]_+\,du,\\
 \log K_{\mathrm e,end}&=rT .
 \end{aligned}
\end{equation}

For known parameters and ideal mean $\mu=0.5$, each MSE is
$(K^2-\mu^2)/10^5$. With calibration included, both methods estimate
the complex endpoint coherence $\widehat h$ on a $+X$ input using
20000 shots per quadrature and reserve $N_{\rm prod}=60000$ independent
shots. Put $\widehat q=\min(1,\max(0.01,|\widehat h|))$ and
$\widehat r=(-\log\widehat q-f_L)/T$. The terminal normalization is
$\widehat q^{-1}$. For the layer normalization we use
\Cref{eq:echo-local-cost} with increments
$\widehat r(t_j-t_{j-1})+f_j-f_{j-1}$. Writing $\widehat K$ for the
respective normalization, the conditional MSE is
\[
 \left(\frac{m}{\widehat q}-\mu\right)^2+
 \frac{\widehat K^2-(m/\widehat q)^2}{N_{\rm prod}},\qquad
 m=\mu\,\frac{\operatorname{Re}(h\widehat h^*)}{|\widehat h|}.
\]
Here $h$ is the true endpoint coherence. At $\widehat h=0$ we choose
a uniform readout phase independently on each shot, giving $m=0$.
Averaging over the two calibration counts gives the MSEs reported in
Table~\ref{tab:background-echo}.

\bibliographystyle{apsrev4-2}
\bibliography{references}
\end{document}